\documentclass{aa}

\usepackage{graphicx}
\usepackage{txfonts}
\usepackage[colorlinks,linkcolor=blue,citecolor=blue,urlcolor=blue,bookmarks=false,hypertexnames=true]{hyperref}
\usepackage{natbib}
\bibpunct{(}{)}{;}{a}{}{,}

\begin{document} 

   \title{Formation of wide-orbit giant planets in protoplanetary disks with a decreasing pebble flux
    }

   \author{Nerea Gurrutxaga
          \inst{1,2},
          Anders Johansen\inst{3,1}, 
          Michiel Lambrechts\inst{3,1}
          \and
          Johan Appelgren\inst{1}
    }

   \institute{Lund Observatory, Division of Astrophysics, Department of Physics, Lund University, Box 43, SE-221 00 Lund, Sweden
         \and
    Max Planck Institute for Solar System Research, Justus-von-Liebig-Weg 3, 37077 Göttingen, Germany\\
         \email{gurrutxaga@mps.mpg.de}
        \and
           Center for Star and Planet Formation, Globe Institute, University of Copenhagen, Øster Voldgade 5-7, 1350 Copenhagen, Denmark\\
    }

   \date{Received 19 September 2023 / Accepted 6 November 2023}

  \abstract
  {The presence of distant protoplanets may explain the observed gaps in the dust emission of protoplanetary disks. Here, we derive a novel analytical model to describe the temporal decay of the pebble flux through a protoplanetary disk as the result of radial drift. This has allowed us to investigate the growth and migration of distant protoplanets throughout the lifespan of the disk. We find that Moon-mass protoplanets that formed early on can grow to their pebble isolation mass, between approximately $20$ and $80\,M_{\oplus}$, within less than $1\,\mathrm{Myr,}$ in the $20$ to $80\,\mathrm{AU}$ region around solar-like stars. The subsequent fast migration in the early stages of gas accretion, after pebble accretion ends, transports these giant planets into their final orbits at $<\,$$10\,\mathrm{AU}$. However, our pebble decay model allows us to include a new pathway that may trigger the transition from pebble accretion to gas accretion after the pebble flux has decayed substantially. With this pebble decay pathway, we show that it is also possible to form gas giants  beyond $10\,\mathrm{AU}$. The occurrence of these wide-orbit gas giants should be relatively low, since their core must attain sufficient mass to accrete gas before the pebble flux decays, while avoiding excessive migration. Since these gas giants do not reach the pebble isolation mass, their heavy element content is typically less than $10\,M_{\oplus}$. Our results imply that the observed gaps in protoplanetary disks could be caused by distant protoplanets that reached the pebble isolation mass and then migrated, while gas giants in wide orbits, such as PDS 70 b and c, accreted their gas after the decay in the pebble flux.}

   \keywords{accretion, accretion disks -- protoplanetary disks -- planets and satellites: general -- planets and satellites: formation -- planets and satellites: gaseous planets  }

    \titlerunning{Formation of wide-orbit giant planets}
   \maketitle
    
%
%-------------------------------------------------------------------

\section{Introduction}
Protoplanetary disks serve as the birthplaces of planets. Spatially resolved observations of these disks have the potential to unravel the physical processes that govern planet formation \citep[see the comprehensive review by][]{Bae2022}. The Atacama Large Millimeter/submillimeter Array (ALMA) has made significant strides in this field by detecting numerous protoplanetary disks exhibiting substructures such as gaps, rings, spirals, and cavities \citep[e.g.,][]{Andrews2018, Zhang2018, Long2018}. While the origins of these substructures remain elusive, they could plausibly be the result of planet-disk interactions, thereby revealing  the presence of hidden planets \citep[e.g.,][]{Lin1986, Crida2006, Pinilla2012}. These hypothetical planets would need to possess considerable masses ($\gtrsim\,$$ 10\,M_{\oplus}$) in order to perturb the surrounding gas within the disk, as well as orbiting at large distances ($>\,$$10\,\mathrm{AU}$), in line with the observed substructure locations \citep[we refer to Fig.\,1 by][and references therein]{Lodato2019}. Moreover, some observations reveal the presence of gaps in disks younger than $1\,\mathrm{Myr}$ \citep[e.g.,][]{Sheehan2018}. If these gaps are indeed a result of the gravitational influence of embedded protoplanets, then it suggests that the formation of massive protoplanets occurs during the early stages of disk evolution.
    
There is already strong evidence supporting the notion that some substructures have formed thanks to the presence of protoplanets. In PDS 70, the giant planets PDS 70 b and c were detected within a cavity, orbiting at $\sim\,$$22\,\mathrm{AU}$ and $\sim\,$$35\,\mathrm{AU,}$ respectively \citep{Keppler2018, Haffert2019}. Another more recent evidence comes from the spirals excited by the planet AB Aur b at $\sim\,$$ 93\,\mathrm{AU}$ from its central star \citep{Currie2022}. These detections do not imply that all detected substructures must be due to the presence of a planet. However, it is necessary to study how common wide-orbit planet formation is and whether it can also statistically explain the origin of the substructures for which no direct evidence of protoplanets has been found. To date, the early formation mechanisms of  distant and massive planets remain an open question.

In the classical picture, the core of a gas giant forms by accreting planetesimals from its vicinity \citep{Wetherill1989, Kokuko&Ida1996, Ormel2010b}, which is followed by the accretion of a gaseous envelope before disk dispersal \citep{Safronov1972, Pollack1996}, typically within a few million years \citep{Haisch2001, Soderblom2014, Tychoniec2020}. Nevertheless, the formation of gas giant cores in the outer regions via planetesimally driven scenarios is hindered by long formation times \citep[e.g.,][]{Thommes2003, Ida2004, Bitsch2015b, JohansenBitsch2019, Lorek2022}. Given this limitation concerning planetesimal accretion, a new paradigm known as pebble accretion was proposed to explain the mechanism that increases the growth rates of forming planets \citep{Ormel2010, Lambrechts2012, Johansen2017}. Pebble accretion allows for a faster growth because pebbles ($\sim\,$mm and cm sized particles) drift radially from the outside of the disk toward the center, continuously replenishing the accreting zone. Moreover, the gas in the vicinity of the protoplanet exerts a drag force that drains kinetic energy from the pebbles. However, such drag force would not be sufficient to increase the accreting rate of $100\,\mathrm{km}$-sized planetesimals.

Although pebble accretion is a prospective mechanism for rapid core formation, the formation of wide-orbit planets still faces challenges such as limited mass reservoirs \citep{Ormel2017} and planetary migration \citep{Ward1997, Johansen2019}. The latter occurs due to the gravitational interaction between the gas and the protoplanet, causing an inward migration to the central star. Indeed, that migration may be rapid enough to prevent the retention of gas giants in wide orbits \citep{Coleman2014}. It is therefore necessary to identify the characteristics that can lead to the formation of distant planets via pebble accretion.

Recent studies of wide-orbit planet formation embraced the idea of pressure bumps or rings. These not only prevent the planet from migrating too quickly, but they also accumulate enough solid build-up for growth \citep[e.g.,][]{Morbidelli2020, Chambers2021, Jiang&Ormel}. However, as previously discussed, the substructures observed in disks might be elicited by other planets. In this work, we therefore focus on the formation of the earliest planets and consider disks with monotonic pressure profiles. We analyze the evolution of individual protoplanets located in the outer regions, with initial masses of $M_{0}$$\,=$$\,0.01\,M_{\oplus}$. The feeding material of these bodies is limited by the depletion of the pebble reservoir. Even though detailed numerical calculations have been implemented to estimate the evolution of solids \citep[e.g.,][]{Brauer2008, Birnstiel2010, Birnstiel2012, Stammler2022}, due to the computational cost of these simulations, the joint study of the evolution of protoplanets while considering pebble flux decay is a difficult task. Some authors have dealt with this issue by employing small pebbles and thereby assuming a tightly coupled evolution of solids and gas \citep[e.g.,][]{Liu2019, Johansen2019}, while others have treated the flux as a free parameter \citep[e.g.,][]{Bitsch2019, Lambrechts2019, Ogihara&Hori2020}. In this work, we present a new analytical expression that is an exact solution to the mass conservation equation for pebbles undergoing radial drift in a viscous gas disk. We validate this expression against the more complex computational analysis as (similarly) conducted by \citet{Appelgren2023}. The new analytical model allows us to explore a new pathway for gas accretion, where the protoplanet can start to accrete once the pebble accretion rate has dropped substantially, in addition to the conventional pathway that requires the core to reach the pebble isolation mass \citep{Lambrechts2014}. The aim of this study is to present the new pebble flux model and to implement it to study the formation of distant cores and gas giants.

This work is structured as follows. In Sect.\,\ref{sec:pebbleAccretion}, we introduce the model of pebble accretion for the outer regions, where we include the derivation of the new analytical model for the pebble flux (in Sect.\,\ref{sec:derivation}). In Sect.\,\ref{sec:coreFormation}, we show the furthest possible core formation in different disk models, which we link to the gaps observed in protoplanetary disks. In Sect.\,\ref{sec:gasAccretion}, we include a simple gas accretion prescription and a new pathway for gas accretion. Furthermore, we highlight the importance of considering pebble depletion to explain gas accretion in distant orbits. We discuss the implications of our findings and the limitations in Sect.\,\ref{sec:implications}. Finally, we summarize our work in Sect.\,\ref{sec:sum}.

%--------------------------------------------------------------------

\section{Pebble accretion in the outer regions}\label{sec:pebbleAccretion}
In this section, we present the model used to elucidate the evolution of a protoplanet in the outer regions. We first describe the disk structure and then proceed to the derivation of a new analytical pebble flux. The outcome of the derivation is specifically given in Eqs.\,(\ref{eq:Z_sol0}) and (\ref{eq:Z_sol1}) under two different assumptions for the pebble Stokes number. Consequently, we describe the growth rates via pebble accretion, followed by the migration of the growing body.

\subsection{Disk structure}

The gas surface density profile of a disk without substructures \citep{Lynden-Bell1974} is described as:
\begin{equation}\label{eq:Sigmag} 
    \Sigma_{\rm{g}} (r, t) = \frac{\dot{\mathcal{M}}_{\rm{g,0}}}{{3}\,\pi\,\nu_{1}\, \tilde{r}^{\gamma}} \, T^{-\frac{5 / 2-\gamma}{2-\gamma}}\,\exp{\left( \frac{-\tilde{r}^{(2-\gamma)}}{T}\right)}\,,
\end{equation}
where $\dot{\mathcal{M}}_{\rm{g,0}}$ is the initial gas accretion rate onto the star in the inner regions, $\nu_{1}$ is the viscosity at the characteristic disk size $R_{1}$, $\gamma$ is the viscous power-law index, $\tilde{r}\,$$\equiv\,$$ r/R_{1}$ is the dimensionless position, and $T$ is the dimensionless time defined through
\begin{equation}\label{eq:T}
    T \equiv \frac{t-t_{0}}{t_{s}} + 1\,.
\end{equation}
Here, $t_{0}$ is the initial time of the disk when $\dot{\mathcal{M}}_{\rm{g}}\,$$=\,$$\dot{\mathcal{M}}_{\rm{g,0}}$, and $t_{\rm{s}}$ is the viscous timescale of the gas at a radial distance of $R_{1}$, which characterizes the time span required for the gas to undergo substantial radial transport. The viscous timescale is defined as:
\begin{equation}\label{eq:ts}
    t_{\rm{s}} \equiv \frac{1}{3(2-\gamma)^{2}} \frac{R^{2}_{1}}{\nu_{1}}\,.
\end{equation}

We describe the viscosity, $\nu,$ using the $\alpha$-disk model \citep{Shakura1973},
\begin{equation} \label{eq:visc}
    \nu = \alpha c_{\rm{s}} H\,, 
\end{equation}
where $c_{\rm{s}}$ is the sound speed and $H$ is the gas scale-height. These two quantities are defined as:
\begin{spacing}{0.3}
\begin{equation}\label{eq:cs}
    c_{\rm{s}}= c_{\rm{s},1} \left( \frac{r}{\mathrm{AU}}\right)^{-\frac{\zeta}{2}}\,,
\end{equation}
\end{spacing}
\begin{equation} \label{eq:H2}
    H = \frac{c_{\rm{s}}}{\Omega}\,,
\end{equation}
where $c_{\rm{s},1}$ is the sound speed at $1\,\mathrm{AU}$, $\zeta$ is the negative power-law index of the temperature, and $\Omega\,$$ =$$ \sqrt{ G M_{\star}/r^{3}}$ is the Keplerian frequency. Equation \ref{eq:visc} relates $\gamma$ and $\zeta$ such that $\gamma\,$$ = \,$$3/2 - \zeta$.

To set the fiducial values, we assume that for a solar-mass star at an initial time of $t_{0}\,$$=\,$$0.2\,\mathrm{Myr}$, there is an accretion rate of \mbox{$\dot{\mathcal{M}}_{\rm{g,0}}\,$$\approx\,$$10^{-7}\,M_{\odot}\,\mathrm{yr^{-1}}$} and an accretion coefficient of $\alpha\,$$\sim\,$$0.01$ \citep{Hartmann1998}. We choose $c_{\rm{s,1}}\,$$= \,$$650\,\mathrm{m\,s^{-1}}$ at $1\,\mathrm{AU}$ \citep{Johansen2019}. In the outer regions where viscous heating can be ignored, $\gamma \,$$\approx\,$$ 15/14$ and $\zeta \,$$\approx\,$$ 3/7$ \citep{Ida2016}. Even though most observed protoplanetary disks appear to be small \citep[e.g.,][]{Barenfeld2017, Tobin2020}, we are interested in understanding substructure formation far from the central star, which calls for large disks. Hence, we assumed a fiducial initial disk size of $R_{1}\,$$ =\,$$ 100\,\mathrm{AU}$. Employing these values, we compute the total mass of the disk as:
\begin{equation}\label{eq:Mdisk} 
\begin{split}
    M_{\rm{g}}(t) =& \int_{0}^{\infty} 2\pi r\Sigma_{\rm{g}}(r, t)\,dr = \frac{2}{3}\frac{\dot{\mathcal{M}}_{g, 0}}{\nu_{1}}\frac{R^{2}_{1}}{(2-\gamma)}T^{\,-\frac{1}{2(2-\gamma)}} \\
    \approx & \;0.17 \,M_{\odot}\; \left( \frac{\alpha}{0.01}\right)^{-1}  \left( \frac{\dot{\mathcal{M}}_{\rm{g,0}}}{10^{-7}\;M_{\odot}\,\mathrm{yr^{-1}}}\right) \\
    & \times \left( \frac{c_{\rm{s,1}}}{650\;\mathrm{m\,\mathrm{s^{-1}}}}\right)^{-2}  \left( \frac{M_{\star}}{M_{\odot}}\right)^{\frac{1}{2}} \left( \frac{R_{1}}{100\,\mathrm{AU}}\right)^{2-\gamma} T^{\,-\frac{1}{2(2-\gamma)}}\,.
\end{split}
\end{equation}
Hence, the initial disk-mass for our fiducial values is \mbox{$M_{\rm{g},0}\,$$\approx\,$$0.17\,M_{\odot}$}. For the same values, but while changing the disk size to a larger disk of $R_{1}\,$$=\,$$300\,\mathrm{AU}$, we instead obtain  $M_{\rm{g},0}\,$$\approx\,$$0.48\,M_{\odot}$. These disks are stable since the Toomre parameter, $Q,$ is higher than 1 everywhere in the disks \citep{Toomre1964}. We list the fiducial values of the parameters in Table \ref{tab:fiducial}.

The radial and temporal dependence of the gas flux is:
\begin{equation}\label{eq:Mdot}
        \dot{\mathcal{M}}_{\rm{g}}(r, t)= \dot{\mathcal{M}}_{\rm{g,0}}\,T^{-\frac{5 / 2-\gamma}{2-\gamma}} \exp \left(-\frac{\tilde{r}^{(2-\gamma)}}{T}\right)\times\left[1-2(2-\gamma) \frac{\tilde{r}^{(2-\gamma)}}{T}\right]\,.
\end{equation}
This flux changes direction at $R_{\rm{t}}\,$$=\,$$ R_{1}\left[ \frac{T}{2(2-\gamma)}\right]^{1/(2-\gamma)}$. By definition, the gas flux is also
\begin{equation}\label{eq:Mgflux}
    \dot{\mathcal{M}}_{\rm{g}} \equiv - 2\pi r v_{\rm{r,g}}  \Sigma_{\rm{g}}\,,
\end{equation}
where $v_{\rm{r,g}}$ is the radial velocity of the gas. The negative sign is due to the definition of a positive flux as being directed toward the star. Knowing $\Sigma_{\rm{g}}$ and $\dot{\mathcal{M}}_{\rm{g}}$, the radial velocity of the gas is:
\begin{equation}\label{eq:vgas}
    v_{\mathrm{r}, \mathrm{g}}= -\frac{\dot{\mathcal{M}}_{\rm{g}}}{2\pi r \Sigma_{\rm{g}}}= -\frac{3}{2} \frac{\nu}{r}\times\left[1-2(2-\gamma) \frac{\tilde{r}^{(2-\gamma)}}{T}\right]\,.
\end{equation}
\begin{figure}
    %\centering
    \includegraphics[width=9cm]{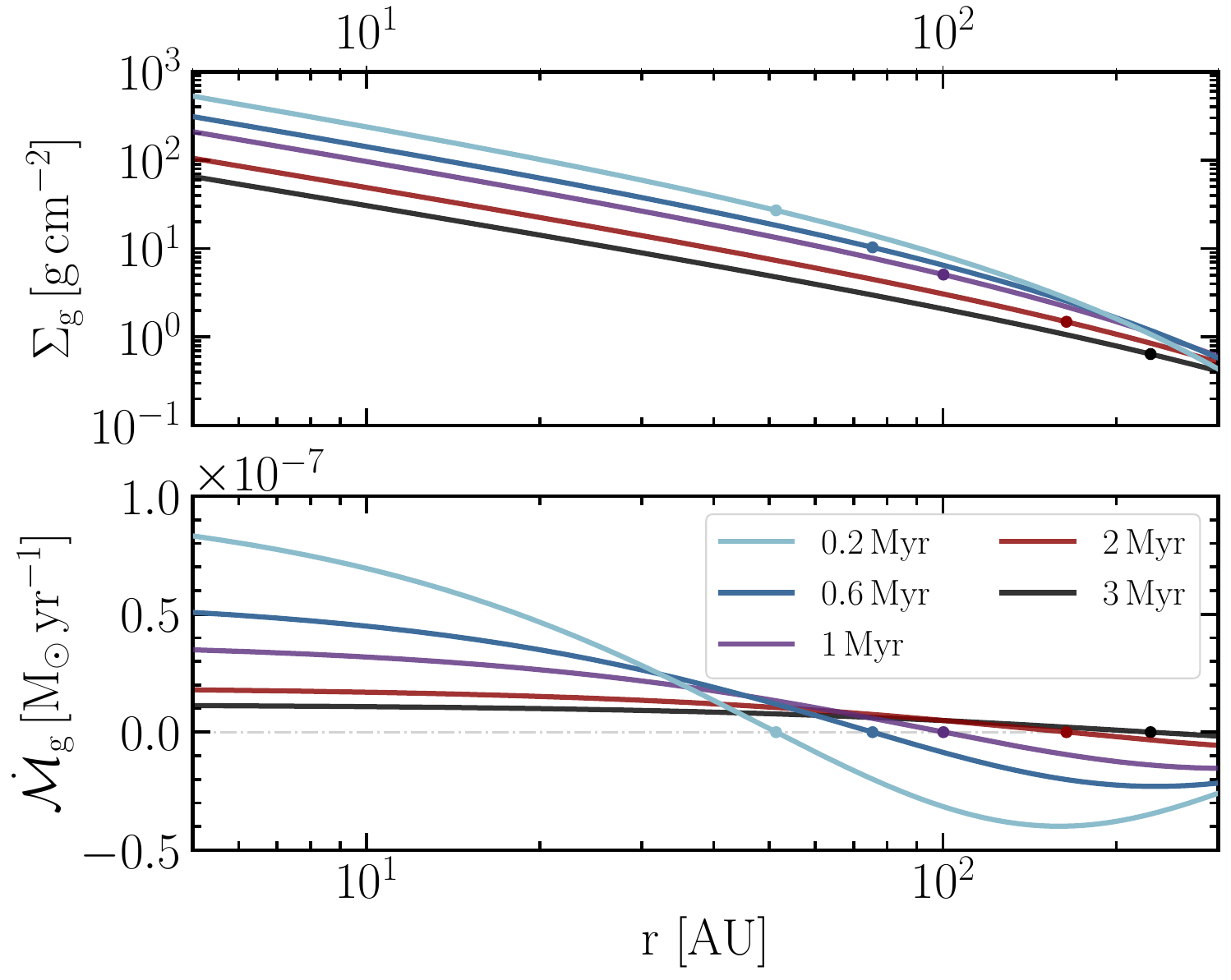}
    \caption{Gas structure in the outer regions of the protoplanetary disk at different times, for our fiducial model with an initial disk size of \mbox{$R_{1}\,$$ =\,$$100\,\mathrm{AU}$} and an initial mass of $0.17\,M_{\odot}$. \textit{Top}: Gas surface density profile from Eq.\,(\ref{eq:Sigmag}). \textit{Bottom}: Radial gas flux from Eq.\,(\ref{eq:Mdot}). The dots illustrate the location where the gas moves outwards with time.}
     \label{fig:density&flux}
     % Plot made by density.py
\end{figure}

In Fig.\,\ref{fig:density&flux}, we show the gas surface density and the gas flux described by Eqs.\,(\ref{eq:Sigmag}) and (\ref{eq:Mdot}), respectively. When the initial disk size is set to $R_{1}\,$$ =\,$$ 100\,\mathrm{AU}$, the outward flux initially emerges at a radius greater than $50\,\mathrm{AU}$ and it then gradually moves outward,
reaching $100\,\mathrm{AU}$ in $1\,\mathrm{Myr}$. Considering the momentum redistribution of the gas is therefore clearly relevant for protoplanets forming at early stages and large distances.

Regarding the orbital motion, the gas is rotating at sub-Keplerian velocities, such that $v_{\phi,\rm{ g}}\,$$ \approx\,$$ v_{\rm{K}} - \Delta v$, where $\Delta v$ is: 
\begin{equation}\label{eq:subKep}
    \Delta v=-\frac{1}{2}\left(\frac{H}{r}\right) \frac{\partial \ln P}{\partial \ln r} c_{\rm{s}} = \frac{1}{2}\left(\frac{H}{r}\right) \chi c_{\rm{s}}\,.
\end{equation}
We denote the negative logarithmic pressure gradient in the midplane as $\chi$,
\begin{equation}\label{eq:chi}
 \chi \equiv -\frac{\partial \ln P}{\partial \ln r}   = \gamma + \frac{\zeta}{2} + \frac{3}{2}  + (2-\gamma)\frac{\tilde{r}^{\,(2-\gamma)}}{T} = \chi_{0} + (2-\gamma)\frac{\tilde{r}^{\,(2-\gamma)}}{T}\,.
\end{equation} 
Here, $\chi_0 \,$$\approx \,$$2.79$ is the pressure gradient in the inner regions of the protoplanetary disk. This pressure gradient is increased in the outer disk by the last term of Eq.\,(\ref{eq:chi}).
    \begin{figure}
        \centering
        \includegraphics[width=9.1cm]{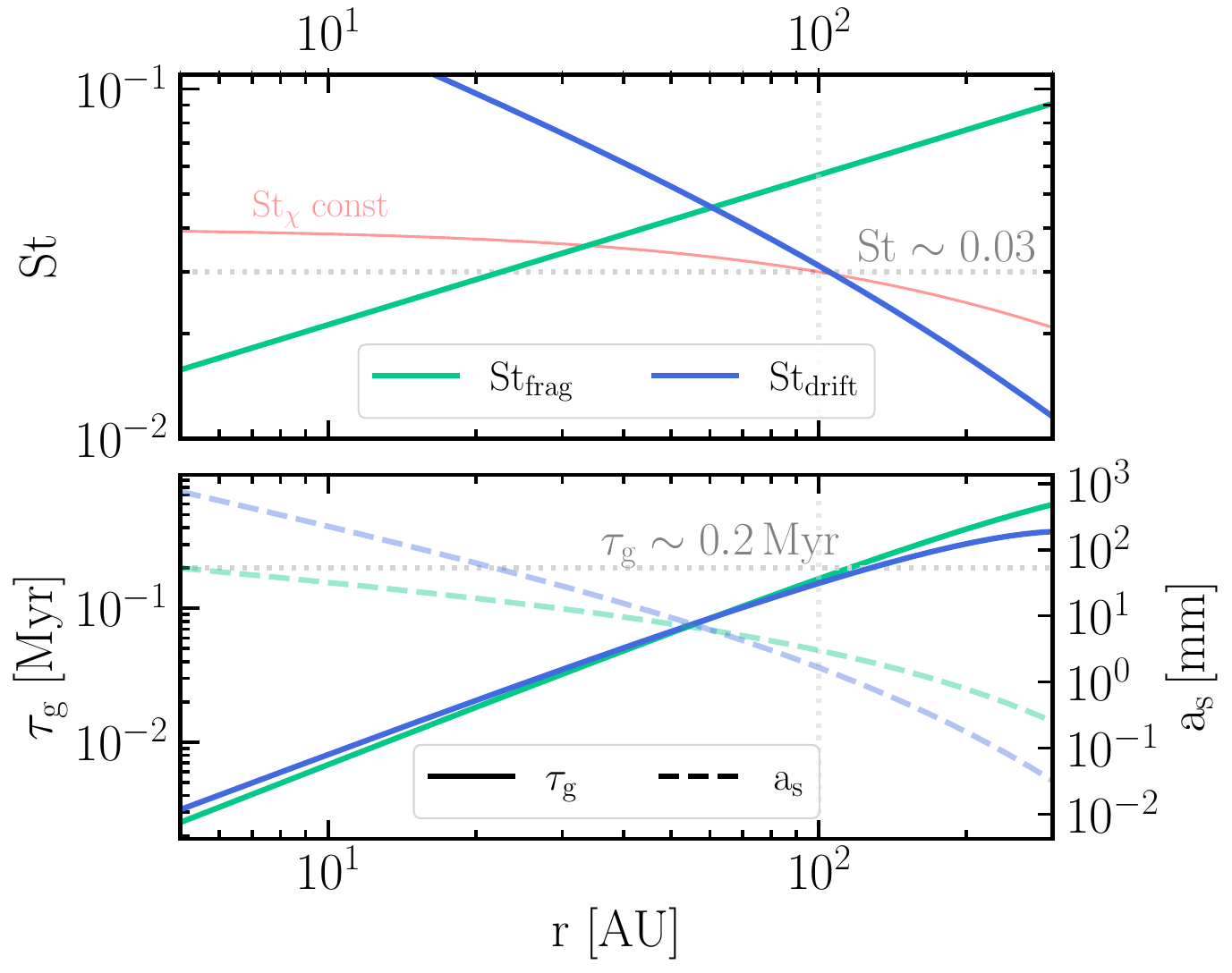}
        \caption{Limiting solid growth barriers across the protoplanetary disk for our fiducial model. Stokes number limited by fragmentation from Eq.\,(\ref{eq:St_fr}), indicated by the green line, and limited by the radial drift from Eq.\,(\ref{eq:St_dr}), indicated by the blue line (top). The red dashed-dotted line illustrates initial $\rm{St}$ when we set constant $\rm{St}_{\chi}\equiv \rm{St}\cdot \chi$ (see Sect.\,\ref{sec:derivation}). Initial particle size for each growth barrier from Eq.\,(\ref{eq:St_a}), indicated by the dashed lines, and the growth timescale for reaching the corresponding size from Eq.\,(\ref{eq:growth}), indicated by the solid lines (bottom). The dotted gray lines denote $\rm{St}\,$$\approx \,$$0.03$ and $\tau_{\rm{g}}\,$$\approx\,$$ 0.2\,\mathrm{Myr}$ at $R_{1}\,$$ =\,$$100\,\mathrm{AU}$ set by radial drift.}
         \label{fig:St_drift}
         % Plot made by St_limit.py
   \end{figure}

\begin{table}
\centering

\caption{Disk parameters employed in this work. Fiducial values are evaluated at position $R_{1}$ and time $t_{0}$ unless indicated.}
\label{tab:fiducial}
\begin{tabular}{ccc}
\hline\hline
\multicolumn{1}{c}{Symbol} & \multicolumn{1}{c}{Definition} & \multicolumn{1}{c}{Fiducial Values}   \\ \hline
$t_{0}$            & Initial time of the disk             & $0.2\,\mathrm{Myr}$\\
$t_{\rm{f}}$            & Disk lifetime             & $5\,\mathrm{Myr}$        \\
$M_{\star}$           & Star mass                     & $1\,M_{\odot}$  \\
$M_{\rm{g},0}$           & Initial disk mass                     & $0.17\,M_{\odot}$    \\
$M_{\rm{s},0}$           & Initial solid mass                     & $570\,M_{\oplus}$        \\
$Z_{0}$           & Initial metallicity                     & $0.01$                       \\
$R_{1}$           & Disk size                     & $100\,\mathrm{AU}$             \\
$\dot{\mathcal{M}}_{\rm{g,0}}$        & Inner gas accretion rate     & $10^{-7}\,M_{\odot}\mathrm{\,yr^{-1}}$             \\
$c_{\rm{s}}$           & Sound speed                     & $650\,\mathrm{m\,s^{-1}}$ at $1\,\mathrm{AU}$                       \\
$\gamma$           & Viscosity power-index                     & $15/14$  \\
$\zeta$           & Temperature power-index                     & $3/7$    \\
$\alpha$           & Accretion viscosity coefficient                     & $0.01$        \\
$\alpha_{\rm{t}}$           & Midplane turbulence                  & $10^{-4}$     \\
$\chi$           & Related to pressure gradient               & $3.71$    \\
$\rm{St}$           & Pebble Stokes number                  & $0.03$  \\
$\rm{St}_{\chi}$           & $\chi\cdot \rm{St}$                  & $0.11$  \\
$v_{\rm{f}}$           & Fragmentation velocity                  & $1\,\mathrm{m\,s^{-1}}$ \\
$\rho_{\rm{s}}$ &   Internal density of particles   & $1\,\mathrm{g\,cm^{-3}}$\\
$\tau_{\rm{g}}$           & Solid growth timescale                & $0.2\,\mathrm{Myr}$ \\
$\kappa$           & Opacity of the envelope                 & $0.005\,\mathrm{~m}^2 \mathrm{~kg}^{-1}$ \\
$\tau_{\rm{th}}$    &  Threshold time   & $10\,\mathrm{Myr}$    \\
$M_{0}$           & Protoplanet's initial mass                  & $0.01\,M_{\oplus}$\\
$r_{0}$           & Protoplanet's initial position                 & $30\rm{-}100\,\mathrm{AU}$ \\
$t_{0,\rm{p}}$           & Protoplanet's formation time                & $0.2-1.2\,\mathrm{Myr}$ \\
\hline

\multicolumn{1}{l}{}         & \multicolumn{1}{l}{}            & \multicolumn{1}{l}{}                            
\end{tabular}
\end{table}
\subsection{Evolution of pebbles}\label{sec:dust}

Since the solid particles initially orbit at Keplerian speed and the gas at sub-Keplerian one, the gas exerts a drag force on the solids, which makes them drift radially in an efficient way when they are pebble-size bodies. The radial velocity of pebbles is described as \citep{Weidenschilling1977}:
\begin{equation}\label{eq:vp}
v_{\mathrm{r}, \rm{p}}=\frac{v_{\mathrm{r}, \mathrm{g}}}{1+\mathrm{St}^2} - \frac{2 \Delta v \,\mathrm{St}}{1+\mathrm{St}^2}\,.
\end{equation}
Here, $\rm{St}$ is the Stokes number, $v_{\rm{r, g}}$ is the radial velocity of the gas from Eq.\,(\ref{eq:vgas}) and $\Delta v$ is the Keplerian velocity reduction of the gas from Eq.\,(\ref{eq:subKep}). The first term in Eq.\,(\ref{eq:vp}) describes the advection mode of transport that occurs when the gas flux drags along solid particles, while the second term corresponds to the radial drift toward higher pressure. In the Epstein regime, $\rm{St}$ is defined as \citep{Weidenschilling1977, Drazkowska2022}:
\begin{equation}\label{eq:St_a}
    \rm{St} = \frac{\pi}{2}\frac{a_{\rm{s}} \rho_{\rm{s}}}{\Sigma_{\rm{g}}}\,,
\end{equation}
where $a_{\rm{s}}$ is the particle size and $\rho_{\rm{s}}$ is its internal density. We chose $\rho_{\rm{s}}\,$$=\,$$1\,\mathrm{g\,cm^{-3}}$ as a nominal value. 

Regarding the growth of solids, the $\mu m$-sized primordial dust particles will easily stick together to form larger particles. In order to determine the particle size, we consider for simplicity that solid growth can be limited either by fragmentation or radial drift \citep[for other growth barriers, see review by][]{Testi2014}. To describe how pairwise collisions can result in fragmentation, the fragmentation velocity $v_{\rm{f}}$ is employed. We adhere to $v_{\rm{f}}\,$$\sim\,$$ 1$$\,\mathrm{m\,s^{-1}}$, as reported by laboratory experiments done by \citet{Guttler2010} for silicate grains. The maximum Stokes number is then described by \citep{Ormel2007, Birnstiel2009}:
\begin{equation}\label{eq:St_fr}
    \mathrm{St_{frag}} \approx \frac{1}{3\alpha_{\rm{t}}}\left(\frac{v_{\rm{f}}}{c_{\rm{s}}}\right)^{2}\,,
\end{equation} 
where $\alpha_{\rm{t}}$ is the midplane turbulence. Since weak gas turbulence has been inferred from dust observations in the outer regions, we chose $\alpha_{\rm{t}}\,$$=\,$$10^{-4}$ as a fiducial value\footnote{The real value could be even lower; for instance, \citet{Villenave2022} found that observations of the disk Oph 163131 are consistent with $\alpha_{\rm{t}}\,$$\lesssim\,$$ 10^{-5}$.} \citep[see review by ][and the references therein]{Pinte2022}. The Stokes number of particles limited by the radial drift is \citep{Ida2016}:
\begin{equation}\label{eq:St_dr}
    \mathrm{St_{drift}} \approx  \frac{\sqrt{3\pi}}{80} \frac{v_{\rm{K}}}{\Delta v} Z_{0}\,.
\end{equation}
The maximum pebble-size at a certain location is the minimum between the fragmentation and drift limit. In the top panel of Fig.\,\ref{fig:St_drift}, we show the fragmentation and drift limits for fiducial values across the protoplanetary disk and we note that at $R_{1}$$\,=$$\,100\,\mathrm{AU,}$ growth is limited by radial drift with $\rm{St}$$\,=$$\,0.03$.

The timescale required to grow from $1\,\mu$m to the particle size $a_{\rm{s}}$ is \citep{Takeuchi2005, Brauer2008, Sato2016}:
\begin{equation}\label{eq:growth}
    \tau_{\rm{g}}\approx  \frac{4}{\sqrt{3\pi}}\frac{1}{Z_{0} \Omega} \ln \left( \frac{a_{\rm{s}}}{\mathrm{\mu m}}\right)\,.
\end{equation}
Here $Z_0$ is the initial metallicity of the disk and $\Omega$ is the Keplerian frequency. In the bottom panel of Fig.\,\ref{fig:St_drift}, we show that the dust growth timescale is approximately $0.2\,\mathrm{Myr}$ at $R_{1}$$\,=$$\,100\,\mathrm{AU}$. Therefore, already at $t\,$$\approx\,$$ 0.2\,\mathrm{Myr}$ pebbles have formed and they drift from the outer to the inner regions of the disk, thus a pebble flux, $\dot{\mathcal{M}}_{\rm{p}}$, can be defined to describe the carried mass \citep{Lambrechts&Johansen2014}. The expression of the pebble flux is given by:
\begin{equation}\label{eq:Mp1}
    \dot{\mathcal{M}}_{\rm{p}}\equiv-2 \pi r v_{\mathrm{r}, \mathrm{p}} \Sigma_{\mathrm{p}}\,,
\end{equation}
where $v_{\mathrm{r},\mathrm{p}}$ is the radial velocity of pebbles from Eq.\,(\ref{eq:vp}) and $\Sigma_{\mathrm{p}}$ is the pebble surface density, which we calculate below.

\subsection{Derivation of the analytical pebble flux}\label{sec:derivation}
Since the dust mass distribution is dominated by the mass of the largest particles, especially in the drift-limited regime \citep{Birnstiel2012}, we assume that the pebble-to-gas ratio is equal to the total metallicity. The pebble-to-gas ratio is defined as:
\begin{equation} \label{eq:Z}
    Z = \frac{\Sigma_{\rm{p}}}{\Sigma_{\rm{g}}}\,.
\end{equation}
The pebble flux can be written in terms of the metallicity by dividing Eqs.\,(\ref{eq:Mgflux}) and (\ref{eq:Mp1}),
\begin{equation} \label{eq:Mp_Z}
    \dot{\mathcal{M}}_{\rm{p}} = Z \frac{v_{\rm{r, p}}}{v_{\rm{r, g}}}\dot{\mathcal{M}}_{\rm{g}}\,.
\end{equation}
For simplicity, we define now an auxiliary quantity $h\,$$=\,$$ \frac{v_{\rm{r, p}}}{v_{\rm{r, g}}}\dot{\mathcal{M}}_{\rm{g}}$ so that $\dot{\mathcal{M}}_{\rm{p}}\,$$=\,$$Z h$. From Eqs.\,(\ref{eq:Mdot}), (\ref{eq:vgas}), and (\ref{eq:vp}), we get the ratio of particle speed to gas speed and, thus, $h(r,t)$ as:
\begin{spacing}{0.3}
\begin{equation}\label{eq:v_ratio_full}
\begin{split}
    \frac{v_{\rm{r, p}}}{v_{\rm{r, g}}} &= \frac{1}{1 + \rm{St}^{2}} \left( 1 - \frac{2 \Delta v \rm{St}}{v_{\rm{r, g}}}\right) \\
    & = \frac{1}{1 + \rm{St}^{2}} \left[ 1 + \frac{2}{3}\frac{\chi \rm{St}}{\alpha}\frac{1}{ 1-2(2-\gamma)\frac{\tilde{r}^{\,(2-\gamma)}}{T}}\right]\,,
\end{split}
\end{equation}
\end{spacing}
\begin{equation}
\begin{split}
    h(r, t) =& \frac{v_{\rm{r, p}}}{v_{\rm{r, g}}}\dot{\mathcal{M}}_{\rm{g}}\\
    =& \frac{1}{1 + \rm{St}^{2}} \left[ 1-2(2-\gamma)\frac{\tilde{r}^{\,(2-\gamma)}}{T} + \frac{2}{3}\frac{\chi \rm{St}}{\alpha}\right] \\
    &\times\dot{\mathcal{M}}_{\rm{g,0}}T^{-\frac{5/2-\gamma}{2-\gamma}}\exp \left({-\frac{\tilde{r}^{\,(2-\gamma)}}{T}}\right)\,.
\end{split}
\end{equation}
Here, $\tilde{r}\,$$=\,$$r/R_{1}$ and $T$ (see Eq.\,\ref{eq:T}) are the dimensionless spatial and time variables. As shown in Eq.\,(\ref{eq:chi}), $\chi$ depends on both $r$ and $t$, and generally $\rm{St}$ can also  depend on $r$ and $t$. Defining then: 
\begin{equation}\label{eq:b}
b(r, t) = \frac{2}{3}\frac{\chi(r, t) \,\mathrm{St} (r, t)}{\alpha}\,,
\end{equation}
we can rewrite $h$ as:
\begin{equation}\label{eq:h(r,t)}
\begin{split}
    h(r, t) =& \frac{1}{1 + \rm{St}^{2}} \left[1 + b(r, t) -2(2-\gamma)\frac{\tilde{r}^{\,(2-\gamma)}}{T} \right] \\
    &\times \dot{\mathcal{M}}_{\rm{g,0}}T^{-\frac{5/2-\gamma}{2-\gamma}}\exp \left({-\frac{\tilde{r}^{\,(2-\gamma)}}{T}}\right)\,.
\end{split}
\end{equation}
Since $h(r,t)$ is a known function of the underlying evolution of the $\alpha$-disk and the speed of the pebbles, our goal now is to derive $Z(r,t)$ so that we can get the analytical form of $\dot{\mathcal{M}}_{\rm{p}}$ in Eq.\,(\ref{eq:Mp_Z}). Neglecting the diffusivity of solid particles within the gas, the pebble flux fulfills its own continuity equation as:
\begin{equation}
    r\frac{\partial \Sigma_{\rm{p}} }{\partial t} -\frac{1}{2\pi} \frac{\partial \dot{\mathcal{M}}_{\rm{p}}}{\partial r} = 0\,.
\end{equation}
We can rewrite this equation in terms of $Z$ by replacing \mbox{$\Sigma_{\rm{p}}\,$$ =\,$$ Z \Sigma_{\rm{g}}$} and $\dot{\mathcal{M}}_{\rm{p}}\,$$ = \,$$Z h$ from Eqs.\,(\ref{eq:Z}) and (\ref{eq:Mp_Z}). We make the change of variables from $r$ and $t$ to the dimensionless parameters $\tilde{r}$ and $T,$ respectively, to simplify the equation. Applying the chain rule, we have:\  
\begin{equation}
    \frac{\partial }{\partial t} \equiv \frac{1}{t_{\rm{s}}}\frac{\partial }{\partial T} \equiv 3(2-\gamma)^{2} \frac{\nu_{1}}{R^{2}_{1}}\frac{\partial }{\partial T}\,, \quad \quad  \frac{\partial }{\partial r} \equiv \frac{1}{R_{1}}\frac{\partial }{\partial \tilde{r}}\,. 
\end{equation}
The continuity equation for the pebbles, therefore, is
\begin{equation}
    3(2-\gamma)^{2}\frac{\nu_{1}}{R_{1}}\tilde{r}\frac{\partial (Z\cdot \Sigma_{\rm{g}})}{\partial T} -\frac{1}{2\pi}\frac{1}{R_{1}}\frac{\partial (Z\cdot h)}{\partial \tilde{r}} = 0\,.
\end{equation}
We simplify the equation by multiplying away $R_{1}$ and expanding the partial derivatives to obtain:
\begin{equation}\label{eq:partial1}
    3(2-\gamma)^{2}\nu_{1}\tilde{r}\left(Z\frac{\partial \Sigma_{\rm{g}}}{\partial T} + \Sigma_{\rm{g}}\frac{\partial Z}{\partial T}\right) -\frac{1}{2\pi}\left(Z\frac{\partial h}{\partial \tilde{r}} + h\frac{\partial Z}{\partial \tilde{r}} \right) = 0\,.
\end{equation}
To solve the partial differential equation, we substitute $\Sigma_{\rm{g}}$ and $h$ from Eqs.\,(\ref{eq:Sigmag}) and (\ref{eq:h(r,t)}) and compute their (dimensionless) temporal and spatial derivatives respectively.
The time derivative of $\Sigma_{\rm{g}}$ is
\begin{equation}
    \frac{\partial \Sigma_{\rm{g}}}{\partial T} =  \frac{\dot{\mathcal{M}}_{\rm{g,0}}}{3\pi\nu_{1}\tilde{r}^{\gamma}} T^{-\frac{5/2-\gamma}{2-\gamma}}\exp \left({-\frac{\tilde{r}^{\,(2-\gamma)}}{T}}\right)\frac{1}{T} \left( -\frac{5/2-\gamma}{2-\gamma} + \frac{\tilde{r}^{\,(2-\gamma)}}{T} \right)\,.
\end{equation}
Computing $\frac{\partial h}{\partial \tilde{r}}$ is not trivial and, therefore, we must make an approximation. The largest pebbles in the outer regions do not usually exceed $\rm{St}\,$$\sim\,$$0.1$. Therefore, for simplicity, $\frac{1}{\rm{St}^{2} + 1}\,$$\approx\,$$ 1$. Consequently, the spatial derivative of $h$ is simplified as:
\begin{equation}
\begin{split}
    \frac{\partial h}{\partial \tilde{r}}= & \biggl\{ \frac{\partial b}{\partial \tilde{r}} - (2-\gamma)\frac{\tilde{r}^{\,(1-\gamma)}}{T} \left[1 + b + 2(2-\gamma)\left( 1 - \frac{\tilde{r}^{\,(2-\gamma)}}{T} \right) \right]\biggl\} \\
    &\times \dot{\mathcal{M}}_{\rm{g,0}}T^{-\frac{5/2-\gamma}{2-\gamma}}\exp \left({-\frac{\tilde{r}^{\,(2-\gamma)}}{T}}\right)\,.
    \end{split}
\end{equation}
By replacing the calculated partial derivatives in Eq.\,(\ref{eq:partial1}) and simplifying terms, the general evolution equation for $Z(\tilde{r},T)$ becomes:
   \begin{equation}\label{eq:general}
   \begin{split}
        & \frac{1}{2}\left[ 1 + b -2(2-\gamma)\frac{\tilde{r}^{\,(2-\gamma)}}{T}\right] \frac{\partial Z}{\partial \tilde{r}} -(2-\gamma)^{2}\tilde{r}^{\,(1-\gamma)}  \frac{\partial Z}{\partial T}
        \\ =& \,Z\frac{\tilde{r}^{\,(1-\gamma)}}{T}\left[ \frac{2-\gamma}{2} b -\frac{T}{2\tilde{r}^{\,(1-\gamma)}} \frac{\partial b}{\partial \tilde{r}} \right]\,.
   \end{split}
   \end{equation}
To compute the solution of this partial differential equation (PDE), first we need to obtain the analytical expression of $b(r,t)$ from Eq.\,(\ref{eq:b}). Since the value of $\rm{St}$ likely changes only weakly with distance in the outer regions (see top panel of Fig.\,\ref{fig:St_drift}), we first assume that $\rm{St}$ is constant. Assuming that the initial pebble-to-gas ratio remains equal to the initial dust-to-gas ratio $Z_{0}$ across the disk, that is, $Z(\tilde{r}, 1)\,$$=\,$$Z_{0}$, the expression of $Z$ as solution of Eq.\,(\ref{eq:general}) is:
\begin{equation}\label{eq:Z_sol0}
    \begin{split}
    Z(\tilde{r}, T) = & Z_{0}T^{\frac{1}{2(2-\gamma)} + \frac{ \rm{St}}{3\alpha}}  \\
    & \times  \exp\biggl\{ {-\left[\frac{1}{2(2-\gamma)}\left(2\chi_{0} + \frac{3\alpha}{\rm{St}}\right) + \frac{\tilde{r}^{(2-\gamma)}}{T}\right] \left[T^{\frac{\rm{St}}{3\alpha}}-1\right]}\biggl\}\,.
    \end{split}
\end{equation}
For the derivation of this result, we refer to Appendix \ref{app:sol_0}. Another possible analytical solution for Eq.\,(\ref{eq:general}) comes from assuming that $\rm{St} \cdot \chi$ is constant, as consequently, $b=b_{0}$ is constant (see Eq.\,\ref{eq:b}). The governing PDE can be simplified to:
\begin{figure}
    \centering
    \includegraphics[width=8.5cm]{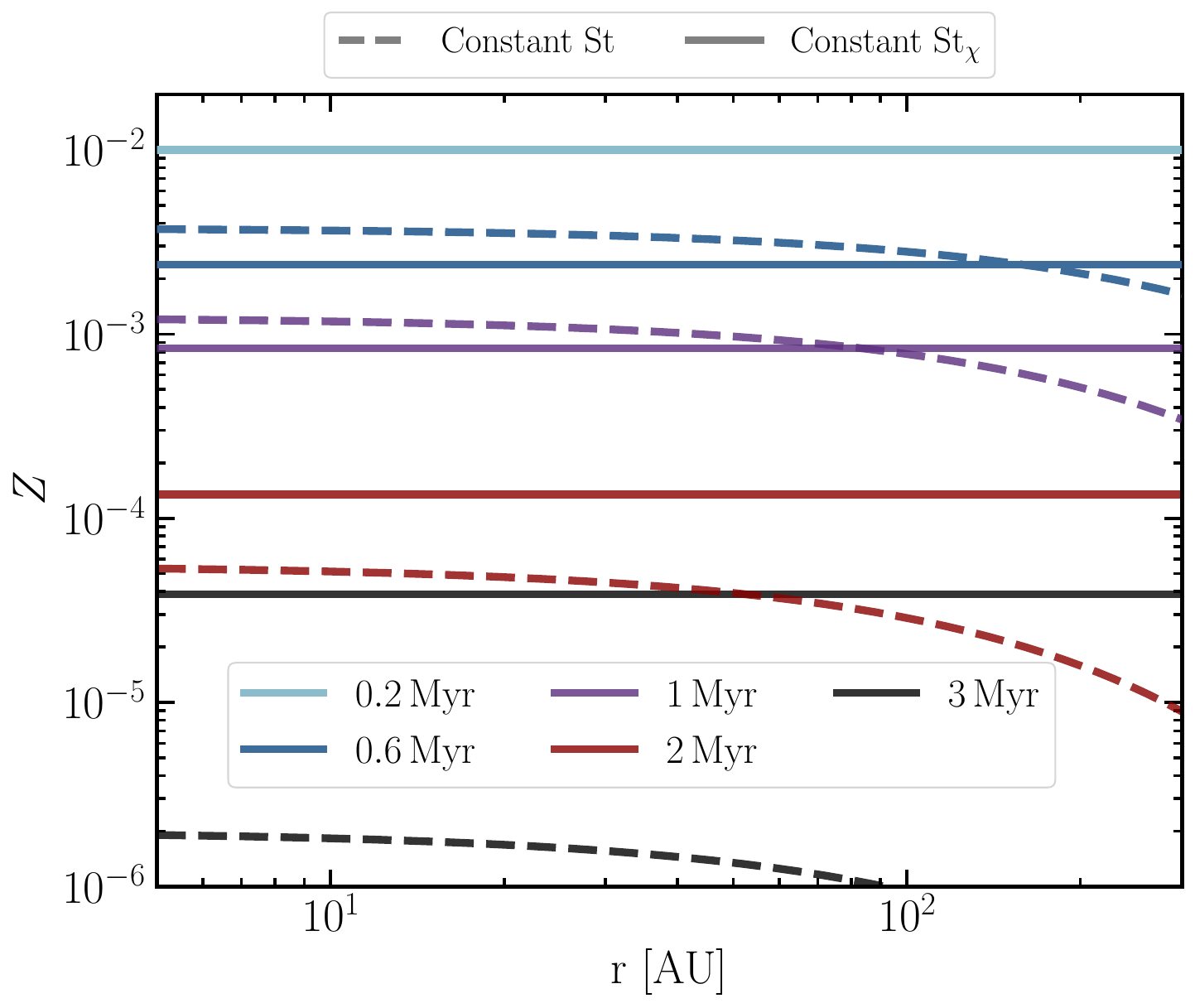}
    \caption{Evolution of the analytical expression for metallicity $Z$ according to constant $\rm{St}$ (Eq.\,\ref{eq:Z_sol0}), indicated by dashed lines, and constant ${\rm St}_\chi$ (Eq.\,\ref{eq:Z_sol1}), indicated by solid lines. For both models, we apply the initial condition $Z_{0}$$\,=$$\,0.01$ at $0.2\,\mathrm{Myr}$ and therefore the initial lines overlap. Then, until the metallicity drops to $\sim$$10\%$ at approximately $1\,\mathrm{Myr}$, both expressions yield a similar evolution. However, the metallicity decreases faster for the constant $\rm{St}$ case in the following $\mathrm{Myrs}$. The employed fiducial values are listed in Table \ref{tab:fiducial}.}
     \label{fig:Z}
     % Plot made by model_testing.py
\end{figure}
\begin{figure*}
    \centering
    \includegraphics[width=17cm]{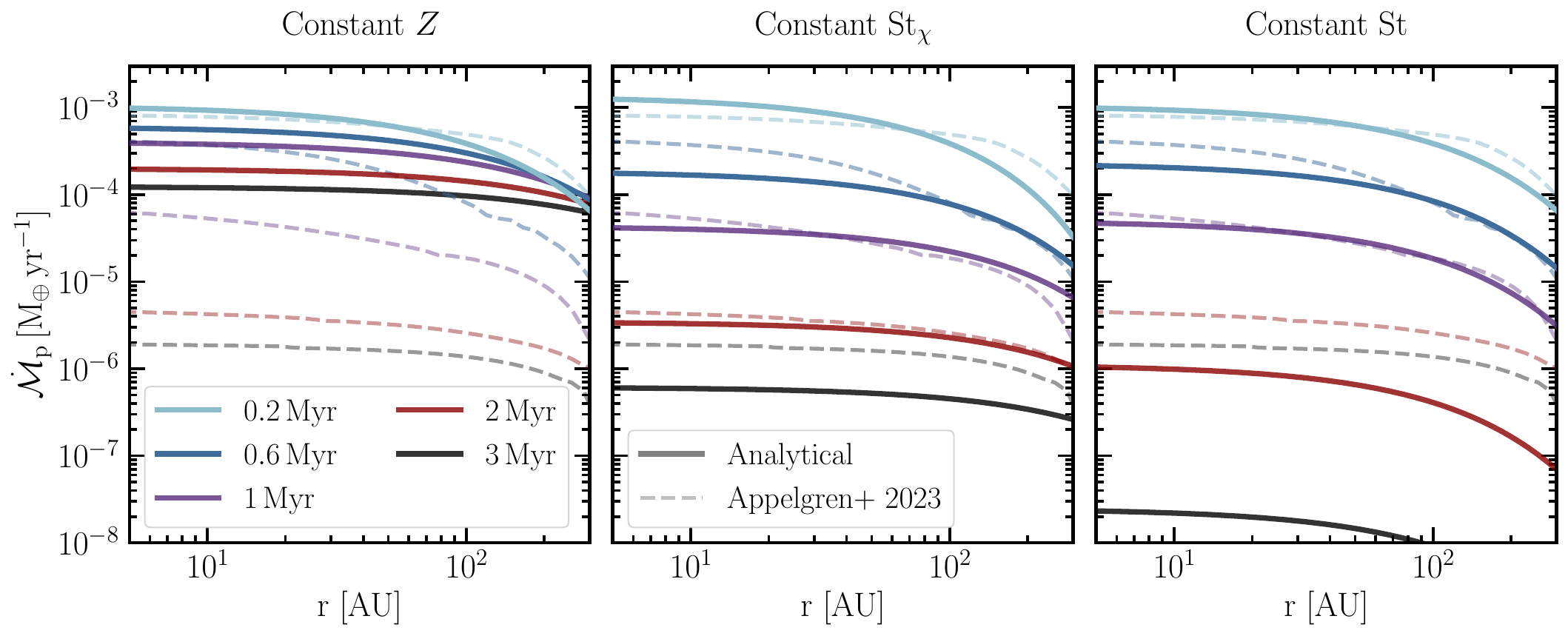}
    \caption{Comparison between the analytical model (solid lines) with the numerical simulations from \citet{Appelgren2023} (dashed lines) at different times. We assume $Z_{0}\,$$=\,$$0.008$ and $\dot{\mathcal{M}}_{\rm{g,0}}\,$$=\,$$6\times 10^{-8}\,M_{\odot}\,\mathrm{yr^{-1}}$ to match the simulation, and $\rm{St}\,$$ =\,$$0.03$ in the three analytical expressions. \textit{Left}: assuming constant metallicity $Z(r, t)\,$$=\,$$Z_{0}$ significantly overestimates the pebble flux over the lifespan of the disk. \textit{Center}: assuming constant $\rm{St}_\chi$ (see Eq.\,\ref{eq:Z_sol0}) gives a relatively good match to the overall behavior of the pebble flux. \textit{Right}: assuming constant $\rm{St}$ gives an accurate description over the first $\sim\,$$ 1\,\mathrm{Myr}$, but it severely underestimates the flux at late times, since the value of $\rm{St}$ is limited by the growth timescale (see Fig.\,\ref{fig:St_drift}) in the outer disk in the simulations.}
     \label{fig:pebbleFlux}
     % Plot made by model_testing.py, dust_drift_rate_no_form.dat
\end{figure*}  
\begin{figure}
    %\centering
    \includegraphics[width=8.5cm]{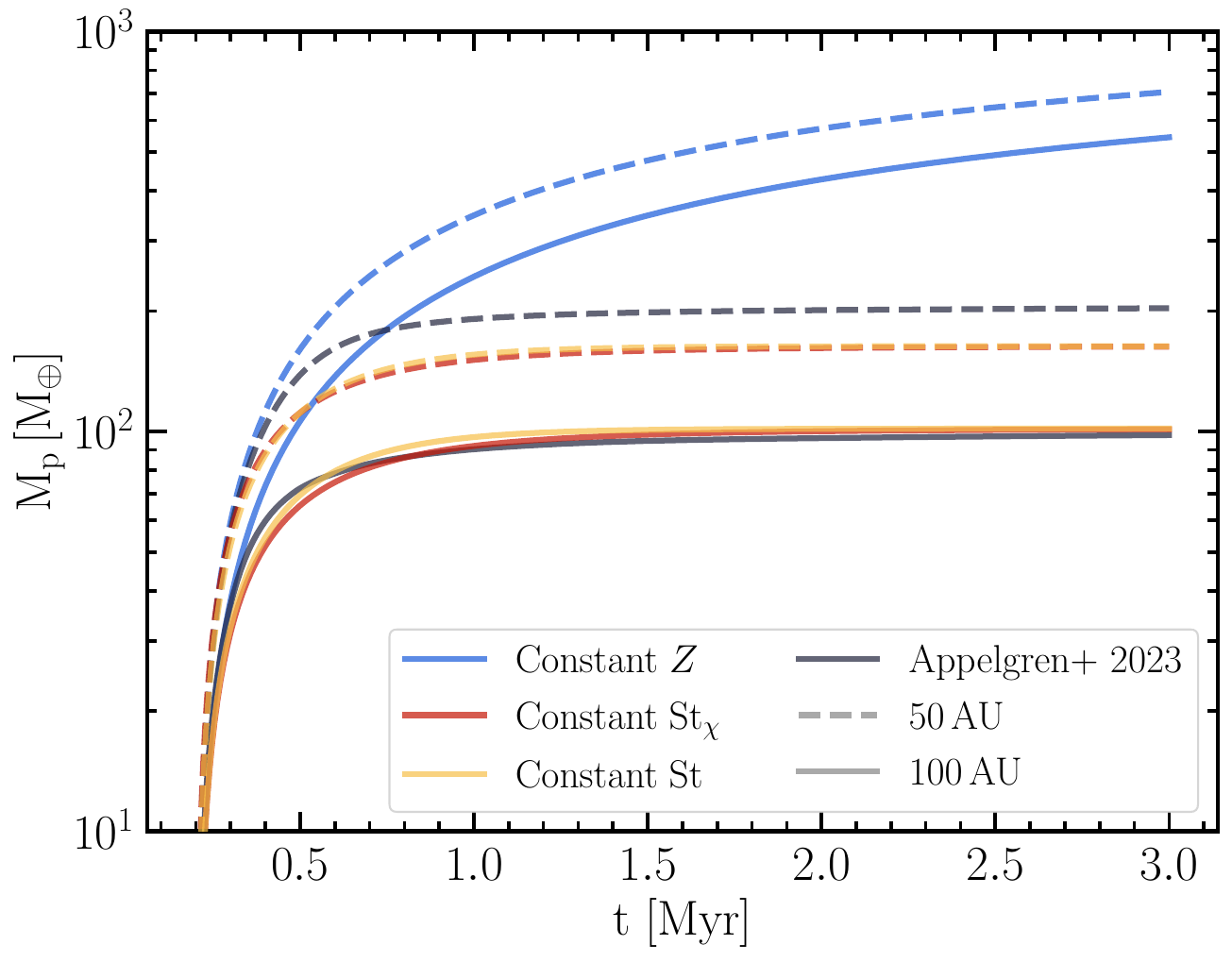}

    \caption{Cumulative mass of drifting pebbles crossing $50\,\mathrm{AU}$ and $100\,\mathrm{AU}$ according to different analytical models and according to the numerical simulation from \citet{Appelgren2023}. In the analytical models shown here, we assume that pebbles have grown to completion and started drifting at $t_{0}\,$$=\,$$0.2\,\mathrm{Myr}$ and that the initial metallicity is $Z_{0}$$\,=$$\,0.008$. Assuming constant $Z$ over time (blue line) significantly overestimates the crossing mass compared to the simulation (black line). In contrast, the cumulative masses from the new analytical models (red and yellow lines) approach the simulated case. We note that at $50\,\mathrm{AU,}$ the cumulative masses from these analytical models are slightly underestimated when comparing to the simulated case since $Z_{0}(50\,\mathrm{AU})\,$$>\,$$0.008$ in the simulation.}
     \label{fig:crossingMass}
     % Plot made by model_testing.py
\end{figure}
\begin{equation}\label{eq:PDE11}
   \begin{split}
        & \frac{1}{2}\left[ 1 + b_{0} -2(2-\gamma)\frac{\tilde{r}^{\,(2-\gamma)}}{T}\right] \frac{\partial Z}{\partial \tilde{r}} -(2-\gamma)^{2}\tilde{r}^{\,(1-\gamma)}\frac{\partial Z}{\partial T} \\
=& \, Z\frac{\tilde{r}^{\,(1-\gamma)}}{T} \frac{2-\gamma}{2} b_{0} \,.
\end{split}
\end{equation}
We introduce an ad hoc assumption to simplify the equation by setting $\frac{\partial Z}{\partial \tilde{r}}\,$$ =\,$$ 0$. Consequently, the PDE reduces to an ordinary differential equation (ODE) since the remaining dependence on $\tilde{r}$ cancels out,
as follows:\begin{equation}
   \begin{split}
        -(2-\gamma)^{2}\tilde{r}^{\,(1-\gamma)}\frac{\partial Z}{\partial T} =& \, Z\frac{\tilde{r}^{\,(1-\gamma)}}{T} \frac{2-\gamma}{2} b_{0} \Rightarrow\\
        -(2-\gamma) \frac{\partial Z}{\partial T} =& \, \frac{Z}{T} \frac{b_{0}}{2}\,.
\end{split}
\end{equation}
Then, we integrate the ODE for the initial conditions $Z(\tilde{r}, 1)\,$$=\,$$Z_{0}$,
\begin{equation}\label{eq:Z_sol1_lagun}
        Z(T)=Z_{0}T^{-\frac{b_{0}}{2(2-\gamma)}}\,.
\end{equation}
The solution is consistent with the assumption $\frac{\partial Z}{\partial \tilde{r}}\,$$=\,$$0$ and,  since the solution of the PDE, which is subject to the specified initial conditions, is unique, Eq.\,(\ref{eq:Z_sol1_lagun}) is, in fact, the solution of the PDE from Eq.\,(\ref{eq:PDE11}). We also include the derivation of Eq.\,(\ref{eq:PDE11}) with the method of characteristics from Appendix \ref{app:sol_1} to show that we get the same result. Since $b_{0}\,$$=\,$$\frac{2}{3}\frac{\chi\cdot\rm{St}}{\alpha}$, the solution becomes:
\begin{equation}\label{eq:Z_sol1}
        Z(T)=Z_{0}T^{-\frac{1}{2-\gamma}\frac{ \chi\cdot\rm{St}}{3\alpha}}\,,
\end{equation}
where $\chi\cdot\rm{St}$ is constant and  is abbreviated as "constant ${\rm St}_\chi$" below.  Later in this paper, we explore how this assumption compares to a model that includes fragmentation limited particle sizes. In order to implement constant $\rm{St}_{\chi}$, we first set a constant of $b_{0}\,$$=\,$$\frac{2}{3}\frac{\chi(R_{1}, t_{0})\cdot \mathrm{St}(R_{1}, t_{0})}{\alpha}$. Then, the variation of $\rm{St}$ in both space and time is described by $\mathrm{St}(r, t)\,$$=\,$$\frac{3}{2}$$\frac{ b_{0}\alpha}{\chi(r, t)}$ with $\chi$ defined in Eq.\,(\ref{eq:chi}), see top panel Fig.\,\ref{fig:St_drift}. In Appendix \ref{app:modelZ_diff}, we demonstrate an alternative derivation of the metallicity, not involving the continuity equation, that yields the same result as in Eq.\,(\ref{eq:Z_sol1}).

To summarize, for constant  values of $\rm{St}$ and $\rm{St}_\chi$, we can derive the evolution of the metallicity via Eqs.\,(\ref{eq:Z_sol0}) and (\ref{eq:Z_sol1}), respectively\footnote{In Appendix \ref{app:St_nonconstant}, we show why we could not solve the PDE analytically when replacing either $\rm{St}_{\rm{drift}}$ nor $\rm{St}_{\rm{frag}}$ in $b$ due to the nonlinearity of the equation.}. In Fig.\,\ref{fig:Z}, we show that the two expressions are similar until the metallicity drops to $\sim$$10\%$ of its original value. After that, the constant St case displays a faster decrease than constant ${\rm St}_\chi$.

The analytical pebble flux is computed by replacing the metallicity in Eq.\,(\ref{eq:Mp_Z}). Even though the full analytical expression of $\dot{\mathcal{M}}_{\rm{p}}$ is nontrivial, we note from Eq.\,(\ref{eq:v_ratio_full}) that at $r$$\,=$$\,0$, the pebble-to-gas flux ratio is described by the simple relation \mbox{$\dot{\mathcal{M}}_{\rm{p}}/\dot{\mathcal{M}}_{\rm{g}} $$\,=$$\,\left( 1 + b \right) Z$}. In Fig.\,\ref{fig:pebbleFlux}, we compare the full analytical pebble flux expression with a numerical simulation as conducted by \citet{Appelgren2023}, excluding disk formation and photoevaporation, and employing the same disk temperature profile as in this work. We also include the case where $Z\,$$=\,$$Z_{0}$ does not evolve (for a detailed comparison with alternative analytical approaches from the literature, see Appendix \ref{app:comparison_dotMp}). Overall, we see that the model with constant $Z\,$$=\,$$Z_{0}$ strongly overestimates the pebble flux after a few hundred thousand years. Otherwise, both of the new analytical expressions imitate the behavior of a more complex computer simulation. After $1\,\mathrm{Myr}$, the constant ${\rm St}_\chi$ case nevertheless describes the pebble flux decay more accurately. In Fig.\,\ref{fig:crossingMass}, we compute the cumulative mass of pebbles at two different locations and, overall, the derived expressions properly estimate the crossing mass.

Given that both constant $\rm{St}$ and constant $\rm{St}_\chi$ assumptions provide an approximated crossing mass value of the flux in the outer regions (and given that a constant $\rm{St}_\chi$ properly describes the decay of the flux), we adhere to the latter to model the planetary growth.

\subsection{Growth via pebble accretion}\label{sec:growth}
The planetesimals in the outer regions of the protoplanetary disk most likely form by the streaming instability (SI) \citep{Johansen2014}. \citet{Lyra2023} found that growth via pebble accretion is possible directly after the planetesimal formation by SI. Thus, we examine here the evolution of a typical initial protoplanet mass of $\sim\,$$0.01\,M_{\oplus}$ that could form directly by SI at large distances \citep{Liu2020}.

To formulate the growth rate via pebble accretion,  we first consider whether the planetary mass is sufficient to accrete from the complete vertical extent of the pebble layer (2D accretion) or not (3D accretion). Secondly, we consider if the relative velocity between the protoplanet and pebbles is dominated by the sub-Keplerian gas flow (Bondi Regime) or by the Keplerian shear (Hill regime).

Regarding the first aspect, the 2D regime is relevant when the pebble accretion radius, $R_{\rm{acc}}$, is larger than the pebble scale-height \citep{Dubrulle1995, Johansen2014}:
\begin{equation}
    H_{\rm{p}} =  H \sqrt{\frac{\alpha_{\rm{t}}}{\alpha_{\rm{t}} + \rm{St}}}\,,
\end{equation}
where $\alpha_{\rm{t}}$ is the aforementioned midplane turbulence. The characteristics favoring the 2D accretion scenario are a large planetary mass and the settling of pebbles. In contrast, in the 3D regime, the protoplanet only has access to a fraction of the pebble layer. Therefore, the efficiency in the 3D regime is lower than in the 2D regime. From \citet{Johansen2017}, the expressions for the growth rate of the protoplanet in each regime are:
\begin{spacing}{0.2}
\begin{equation}\label{eq:M2d}
     \dot{M}_{\rm{2D}} = 2R_{\rm{acc}}\Sigma_{\rm{p}} \delta v\,,
\end{equation}
\end{spacing}
\begin{equation}\label{eq:M3d}
     \dot{M}_{\rm{3D}} = \pi R^{2}_{\rm{acc}}\rho_{\rm{p}} \delta v\,,
\end{equation}
where $\Sigma_{\rm{p}}$ is the pebble surface density that can be analytically derived from Eqs.\,(\ref{eq:Z}) and (\ref{eq:Z_sol1}), 
\begin{equation}\label{eq:Sigmap} 
    \Sigma_{\rm{p}} (r, t) = \frac{\dot{\mathcal{M}}_{\rm{g,0}}}{{3}\,\pi\,\nu_{1}\, \tilde{r}^{\gamma}} Z_{0}\, T^{-\frac{1}{2-\gamma}\left( \frac{5}{2}-\gamma +\frac{\chi\rm{St}}{3\alpha}\right)}\,\exp{\left( \frac{-\tilde{r}^{(2-\gamma)}}{T}\right)}\,,
   \end{equation}
assuming that all dust grows into pebbles. In Eq.\,(\ref{eq:M3d}), \mbox{$\rho_{\rm{p}}\,$$=\,$$\frac{1}{\sqrt{2\pi}} \frac{\Sigma_{\rm{p}}}{H_{\rm{p}}}$} is the pebble density in the midplane. Both rates depend on $\delta v$, the approach velocity between the pebbles and the protoplanet, defined as:
 \begin{equation} \label{eq:approach_v}
 \delta v \equiv \Omega R_{\rm{acc}}+ \Delta v\,,
 \end{equation}
 where $\Delta v$ is the sub-Keplerian velocity reduction of the gas from Eq.\,(\ref{eq:subKep}). For the transition between 3D and 2D to be continuous, $\dot{M}_{\rm{2D}}\,$$=\,$$ \dot{M}_{\rm{3D}}$ must hold at a certain accretion stage. From this equality, we get that the transition occurs when:
 \begin{equation}\label{eq:Racc_Hp}
     \frac{R_{\rm{acc}}}{H_{\rm{p}}} = \sqrt{\frac{8}{\pi}}\approx 1.6\,.
 \end{equation}

\begin{figure}
        \centering
        \includegraphics[width=9cm]{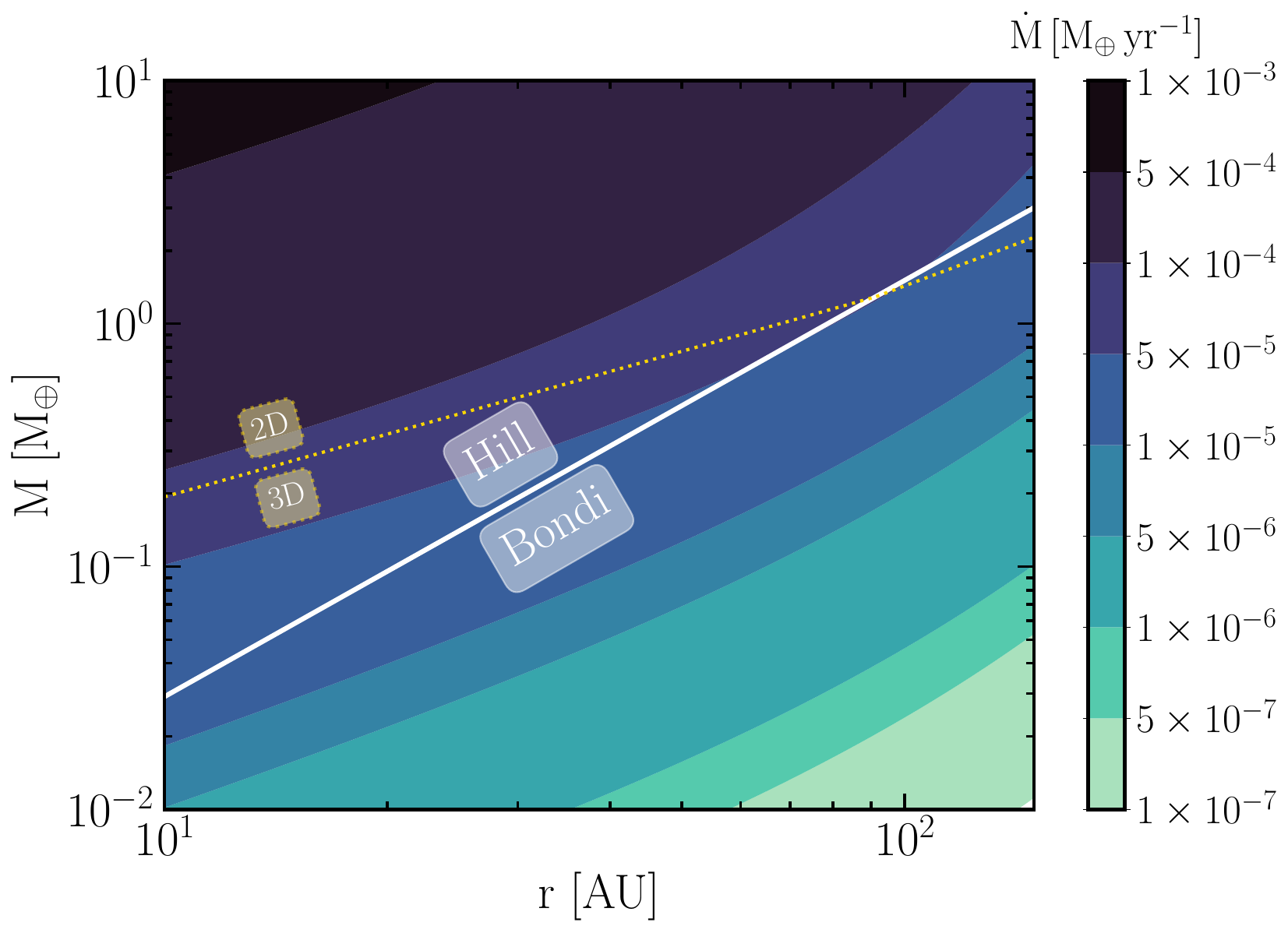}
        \caption{Initial pebble accretion rate onto a protoplanet as a function of its position and mass. The separation between 3D and 2D accretion regimes and between Bondi and Hill regimes are indicated. In the outer regions, we need to consider 3D and 2D as well as Bondi and Hill regimes. Employed fiducial values are listed in Table \ref{tab:fiducial}.}
         \label{fig:BondiHill}
         % Plot made by BondiHill.py
   \end{figure}

The analytical form of the accretion radius $R_{\rm{acc}}$ will depend on whether the protoplanet is accreting in the Bondi or Hill regime. The Hill and Bondi radius of a protoplanet are defined as:
\begin{spacing}{0.3}
\begin{equation}
    R_{\rm{H}} \equiv r \left( \frac{M}{3M_{\star}}\right)^{1/3}\,,
\end{equation}
\end{spacing}
\begin{equation}
    R_{\rm{B}} \equiv \frac{GM}{\Delta v^{2}}\,,
\end{equation}
where $r$ and $M$ are the position and mass of the protoplanet. \citet{Johansen2017} derived an expression for the effective accretion radius in each regime,
as follows:\begin{spacing}{0.3}
\begin{equation} \label{eq:Rh}
    R_{\rm{acc}} =\left( \frac{\rm{St}}{0.1} \right)^{1/3} R_{\rm{H}}\, \quad \quad \quad \;\textnormal{(Hill regime),}
\end{equation}
\end{spacing}
\begin{equation}\label{eq:Rb}
    R_{\rm{acc}} = \left(  \frac{4 \rm{St} \,\Delta v}{\Omega\,R_{\rm{B}}} \right)^{1/2} R_{\rm{B}}\,\quad \quad \textnormal{(Bondi regime),}
\end{equation}
and by equating the two accretion radii, we can determine the transitional mass between the two regimes as: 
\begin{equation}\label{eq:tranMass}
    M_{\rm{t}} = \frac{25}{144} \frac{\Delta v^{3}}{G \Omega}\frac{1}{\rm{St}}\,.
\end{equation}
The accretion will occur in the Hill regime if $M\,$$\geq \,$$M_{\rm{t}}$ and in the Bondi regime if $M\,$$<\,$$ M_{\rm{t}}$. In Fig.\,\ref{fig:BondiHill}, we show the accretion rate calculated via Eqs.\,(\ref{eq:M2d}) or (\ref{eq:M3d}), depending on Eq.\,(\ref{eq:Racc_Hp}), and by replacing Eqs.\,(\ref{eq:Rh}) or (\ref{eq:Rb}), depending on Eq.\,(\ref{eq:tranMass}). This approach ensures a smooth and continuous growth rate during the transition from one regime to another.

   \begin{figure*}
       %\centering
       \includegraphics[width=17cm]{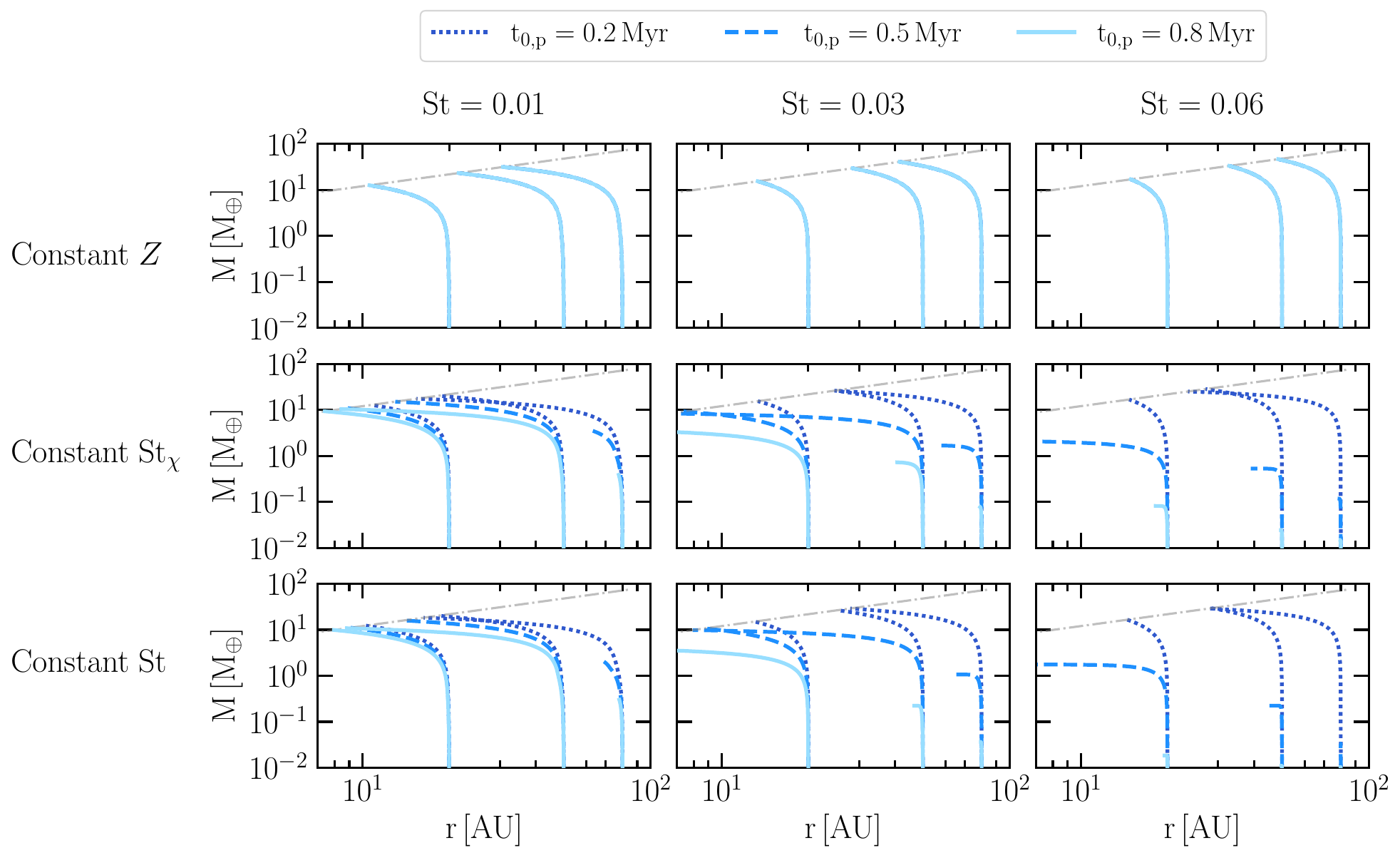}
       \caption{Numerically integrated growth tracks for three different pebble fluxes or metallicity models (row-to-row) and different Stokes numbers (column-to-column). In the constant $\rm{St}_{\chi}$ model, $\rm{St}$$\,=$$\,0.01,0.03$ and $0.06$ are employed to set the initial $\rm{St}$ at radial distance $R_{1}$ (see top panel Fig.\,\ref{fig:St_drift}). Protoplanets with initial mass of $M_{0}\,$$=\,$$0.01\,M_{\oplus}$ are placed at $r_{0}\,$$=\,$$20, 50$ or $80\,\mathrm{AU}$ at different formation times (values indicated in the legend). We stop the integration either at the end of the disk lifetime of $t_{\rm{f}}\,$$ =\,$$5\,\mathrm{Myr}$ or earlier if they reach $M_{\rm{iso}}$, indicated by the gray dashed-dotted line. When $Z$ is constant, which is equivalent to assuming that the pebble-to-gas flux ratio in the inner regions is constant, the growth tracks are independent of $t_{0,\rm{p}}$. Consequently, the growth is overestimated compared to constant $\rm{St}$ and $\rm{St}_{\chi}$, especially when $\rm{St}\,$$ =\,$$0.03$ or $0.06$ and when $t_{0,\rm{p}}\,$$=\,$$ 0.5\,\mathrm{Myr}$ or $0.8\,\mathrm{Myr}$. Models with constant ${\rm St}_\chi$ and $\rm{St}$ from Eqs.\,(\ref{eq:Z_sol1}) and (\ref{eq:Z_sol0}), respectively, yield similar results.}
        \label{fig:ModelComparison}%
        % Plot made by track_comparison.py
    \end{figure*}

The maximum core mass that the protoplanet can attain is known as the pebble isolation mass, $M_{\rm{iso}}$ \citep{Lambrechts2014}. According to the 3D simulations in \citet{Bitsch2018}, if the protoplanet grows massive enough to carve out a gap of a depth of $\sim\,$$ 10-20\%$, the pressure gradient in the outer edge is reversed and pebbles are pushed outward, causing pebble accretion to cease and enabling gas accretion to occur. The authors derived a scaling law of $M_{\rm{iso}}$ for protoplanets orbiting solar-mass stars, such that:
\begin{equation} \label{eq:Miso}
\begin{split}
     M_{\rm{iso}}(r)=& 25 M_{\oplus}\left( \frac{H/r}{0.05}\right)^{3} \times  \left[ 0.34 \left( \frac{\log{10^{-3}}}{\log{\alpha_{\rm{t}}}}\right)^{4} + 0.66\right]  \\
     & \times \left[ 1 - \frac{-\chi_{0} + 2.5}{6}\right]\,,
\end{split}
\end{equation}
where $\alpha_{\rm{t}}$ is the midplane turbulence and $\chi_{0}$ is the negative logarithmic pressure gradient in Eq.\,(\ref{eq:chi}). The pebble isolation mass increases for small pebbles \citep{Bitsch2018}, but we neglected this effect due to our choice of relatively large pebbles.

\subsection{Migration}\label{sec:migration}

Protoplanets undergo inward radial migration while they grow. Planets that are not massive enough to open a gap in the gas fall into the type-I migration regime. We describe the migration speed of these protoplanets by the standard scaling law derived in \citet{Tanaka2002},
\begin{equation} \label{eq:migI}
\dot{r}=-k_{\mathrm{mig}} \frac{M}{M_{\star}} \frac{\Sigma_{\mathrm{g}} r^{2}}{M_{\star}}\left(\frac{H}{r}\right)^{-2} v_{\mathrm{K}}\,,
\end{equation}
where $v_{\mathrm{K}}$ is the Keplerian velocity and $k_{\mathrm{mig}}$ the constant prefactor that was fitted using 3D numerical simulations in \citet{D'Angelo2010}
\begin{equation}
    k_{\rm{mig}} = -2\,(1.36+0.62 \gamma+0.43 \zeta)\,.
\end{equation}
Here, $\gamma$ and $\zeta$ are the already discussed power-law indexes of the viscosity and the midplane temperature or sound speed, respectively.

When a planet reaches a certain mass threshold, $M_{\rm{gap}}$, it creates a density gap along its orbit that leads to reduced migration rates. If the gas does not flow through this gap, the planet is forced to migrate at the same speed as the viscous accretion of the gas \citep{Lin1986}. This type of migration is referred to as type II migration. However, hydrodynamical simulations have demonstrated that gas will cross the gap \citep[e.g.,][]{Durmann2015}. In \citet{Kanagawa2018}, a new physical model was proposed in which the torque exerted by the gas that crosses the gap depends on the surface density at the bottom of the gap, $\Sigma_{\rm{gap}}$, rather than the unperturbed density, $\Sigma_{\rm{g}}$, as in the type-I regime. Then, $\Sigma_{\rm{gap}}$ decreases as the protoplanet's mass increases, thereby slowing down migration. The migration effectively slows down when the gap depth ($\Sigma_{\rm{gap}}/\Sigma_{\rm{g}}$) is reduced to approximately $50\%$. Consequently, the protoplanet reaches $M_{\rm{iso}}$ slightly before $M_{\rm{gap}}$. In \citet{Johansen2019}, it was suggested that a relative gap height of around $85\%$ is sufficient to reach $M_{\rm{iso}}$. They found that $M_{\rm{gap}}\,$$\approx \,$$2.3\,M_{\rm{iso}}$ and provided the modified migration equation as follows:
\begin{equation}
    \dot{r}=\frac{\Sigma_{\mathrm{gap}}}{\Sigma_{\mathrm{g}}}\cdot \dot{r}_{\mathrm{I}}= \frac{\dot{r}_{\mathrm{I}}}{1+\left[M /\left(2.3 M_{\mathrm{iso}}\right)\right]^2}\,,
\end{equation}
where $\dot{r}_{\mathrm{I}}$ represents the classical type-I migration rate from Eq.\,(\ref{eq:migI}) and $M_{\rm{iso}}$ is described in Eq.\,(\ref{eq:Miso}). 

\section{Formation of wide-orbit cores}\label{sec:coreFormation}
In this section, we analyze the evolution of protoplanets that grow via pebble accretion while migrating until they reach the pebble isolation mass, $M_{\rm{iso}}$. We also included gas accretion (see Sect.\,\ref{sec:gasAccretion}).

\begin{figure*}
   \centering
   \includegraphics[width=17cm]{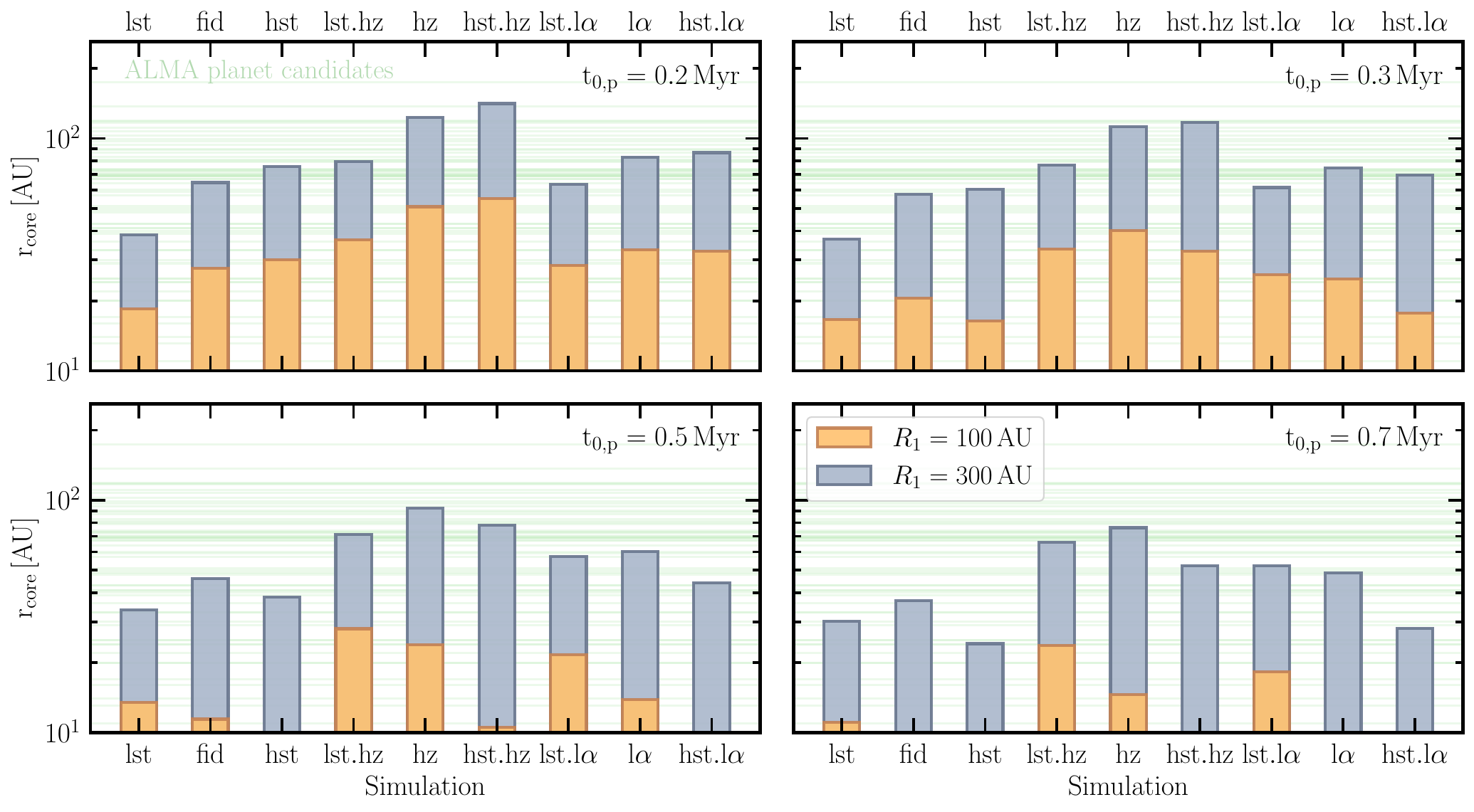}
   \caption{Location of the furthest cores formed by reaching $M_{\rm{iso}}$, marked by the height of the colored bar, for different scenarios (specified in Table \ref{tab:simSet}) and for protoplanets initialized at different formation times, $t_{0,\rm{p}}$, in each panel. Horizontal green lines in the background illustrate the location of the observed gaps in protoplanetary disks that could be caused by planetary cores placed in the same location\protect\footnotemark. When $t_{0,\rm{p}}\,$$ =\,$$0.2\,\mathrm{Myr}$ (top-left panel), favorable disk parameters for forming distant cores up to $\sim\,$$ R_{1}/2$ are high $\rm{St}$, high $Z_{0}$ and low $\alpha_{\rm{t}}$. "hst" runs with $\rm{St}$$\,=$$\,0.06$ also show that when initial protoplanets are injected late ($t_{0,\rm{p}}$$\,\geq$$\,0.5\,\mathrm{Myr}$, bottom panels) in disks with $R_{1}$$\,=$$\,100\,\mathrm{AU}$, the pebble flux by that point is too weak to grow wide-orbit cores. We note that for $R_{1}\,$$ =\,$$300\,\mathrm{AU}$, the $t_{0,\rm{p}}\,$$ =\,$$0.7\,\mathrm{Myr}$ case is more consistent with the pebble growth timescale in outer disk (see Fig.\,\ref{fig:St_drift}). Overall, many observed gaps in protoplanetary disks could be caused by planetary cores, but the most distant gaps could only form in disks with high initial metallicity and a large disk size (see Sect.\,\ref{sec:gaps} for discussion).}
    \label{fig:furthestCore}%
    % Plot made by find_distantcore.py
\end{figure*}
\subsection{Growth tracks with the new pebble flux model}
\footnotetext{Data from \url{http://ppvii.org/chapter/12/figure7.txt/}.}
We applied the new analytical models derived in Sect.\,\ref{sec:derivation} for calculating the growth tracks of protoplanets and compared them with the results obtained by the assumption of constant metallicity (or coupling between solids and gas) over time.

For that purpose, we initialized the protoplanets of \mbox{$M_{0}\,$$=\,$$0.01\,M_{\oplus}$} formed at $t_{\rm{0,p}}\,$$=\,$$0.2,0.5$ or $0.8\,\mathrm{Myr}$ and located at $r_{0}\,$$=\,$$20, 50$ or $80\,\mathrm{AU}$. We placed them within a disk for the three different descriptions of the pebble flux (or metallicity) and we varied the pebble Stokes number as $\rm{St}\,$$ =\,$$0.01, 0.03,$ or $0.06$. For the rest of the disk parameters, we adhered to the values motivated in Sect.\,\ref{sec:pebbleAccretion} and listed in Table \ref{tab:fiducial}. We calculated the planetary growth and migration as described in Sects.\,\ref{sec:growth} and \ref{sec:migration}, respectively, and stopped the calculations either at the end of the disk lifetime (we set $t_{\rm{f}}\,$$ =\,$$5\,\mathrm{Myr}$) or earlier if they have reached the value of $M_{\rm{iso}}$ from Eq.\,(\ref{eq:Miso}).

We show the results of these calculations in Fig.\,\ref{fig:ModelComparison}. For \mbox{$\rm{St}\,$$ =\,$$0.01$}, the growth tracks of protoplanets started at \mbox{$t_{\rm{0,p}}\,$$=\,$$0.2\,\mathrm{Myr}$} are similar in the three models. However, for protoplanets formed at $t_{\rm{0,p}}\,$$=\,$$0.5$ and $0.8\,\mathrm{Myr}$, the growth is overestimated when using the model with a constant $Z$. For $\rm{St}\,$$=\,$$0.03$, the growth in the outer regions is overestimated even for protoplanets formed at $t_{0,\rm{p}}\,$$ =\,$$0.2\,\mathrm{Myr}$. In contrast, models with constant $\rm{St}_{\chi}$ and $\rm{St}$ have similar outcomes. Finally, for $\rm{St}\,$$=\,$$0.06$, the $Z$ constant model only gives proper results when $t_{0,\rm{p}}\,$$ =\,$$0.2\,\mathrm{Myr}$ and $r_{0}\,$$=\,$$20\,\mathrm{AU}$.

Overall, Fig.\,\ref{fig:ModelComparison} demonstrates that it is necessary to consider the decay of the pebble flux to prevent overestimating the growth. For the subsequent calculations, we therefore adhere to the $Z(t)$ model with the constant $\rm{St}_{\chi}$ from Eq.\,(\ref{eq:Z_sol1}).

\begin{table}
    \caption{Disk parameters of simulated scenarios. Abbreviations adhere to the following format: prefix l/h (low and high) + varied parameter $\rm{St}$/$Z_{0}$/$\alpha_{\rm{t}}$ (Stokes number/initial metallicity/midplane turbulence).}
    \label{tab:simSet}
    \centering
    \begin{tabular}{c c c c}
        \hline\hline
        Scenario      &  $\rm{St}$    &   $Z_{0}$   &   $\alpha_{\rm{t}}$ \\
        \hline
        lSt & 0.01 & 0.01 & $10^{-4}$     \\
        fid & 0.03 & 0.01 & $10^{-4}$            \\
        hSt & 0.06 & 0.01 & $10^{-4}$            \\
        lSt.hZ & 0.01 & 0.02 & $10^{-4}$     \\
        hZ & 0.03 & 0.02 & $10^{-4}$            \\
        hSt.hZ & 0.06 & 0.02 & $10^{-4}$            \\
        lSt.l$\alpha$ & 0.01 & 0.01 & $10^{-5}$     \\
        l$\alpha$ & 0.03 & 0.01 & $10^{-5}$            \\
        hSt.l$\alpha$ & 0.06 & 0.01 & $10^{-5}$            \\
    \hline   
\end{tabular}
\end{table}

\subsection{Location of the furthest cores in different scenarios}\label{sec:distantcores}

Given that the protoplanet reaches $M_{\rm{iso}}$ by reversing the pressure gradient when carving out a gap in its vicinity, the formation of the core is related to the formation of gaps. Hence, if we can demonstrate that the formation of distant cores is possible, we can elucidate a plausible origin for the observed gaps far away from their central star. For this purpose, we investigated how far out cores of giant planets can form by reaching $M_{\rm{iso}}$ in different disks. We  then compared the location of the furthest cores with the observed gap locations in protoplanetary disks.

    \begin{figure*}
       \centering
       \includegraphics[width=14.0cm]{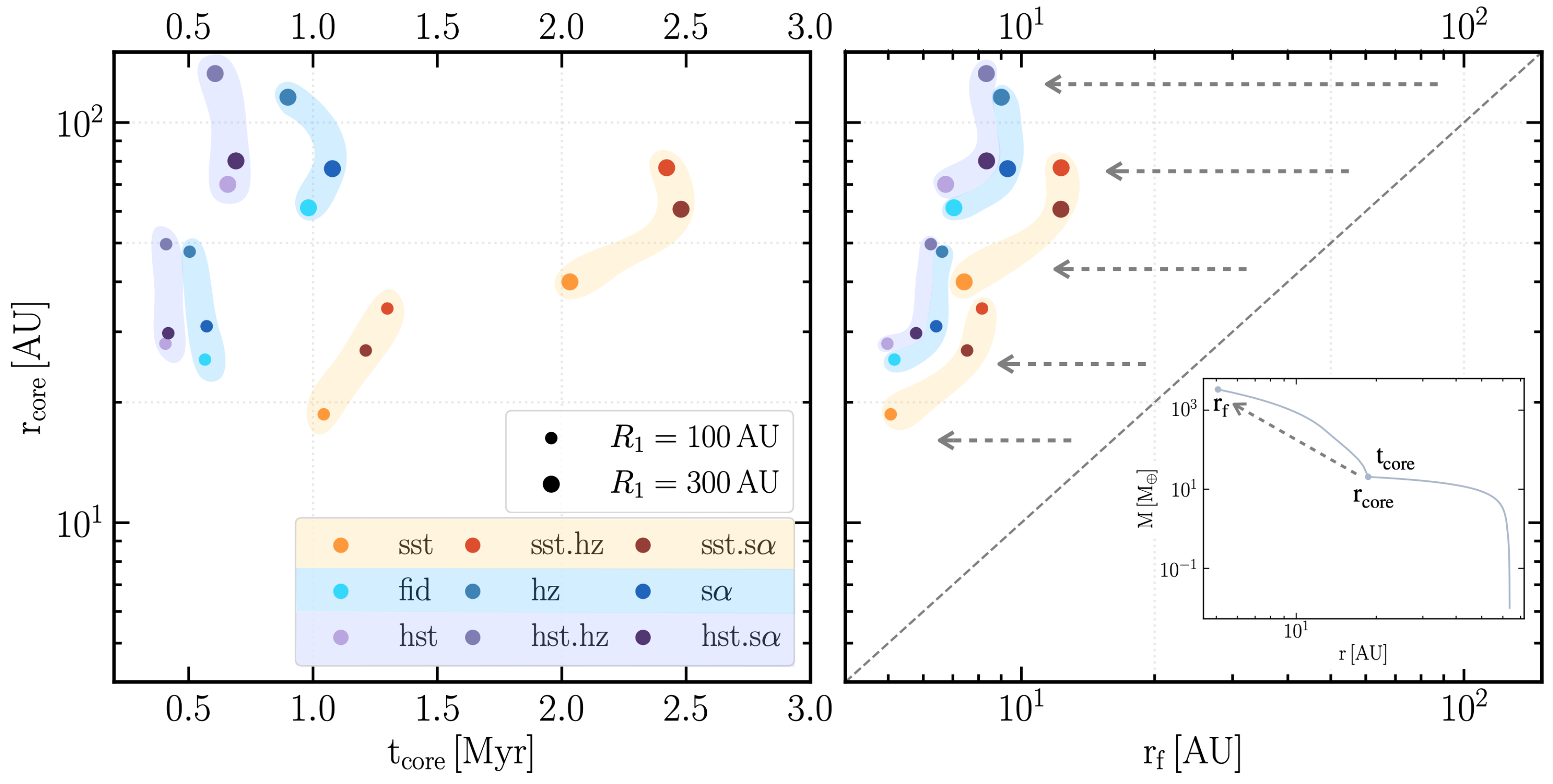}
       \caption{Location and formation time of the furthest cores simulated in different scenarios from Table \ref{tab:simSet} when protoplanets emerge at \mbox{$t_{0,\rm{p}}$$\,=$$\,0.2\,\mathrm{Myr }$} (left). Scenarios with the same $\rm{St}$ are highlighted by their colored groups.  Final location of the furthest cores after gas accretion (Sect.\,\ref{sec:gasAccretion_mini}) and type II migration (Sect.\,\ref{sec:migration}) at the end of the disk lifetime of $5\,\mathrm{Myr}$ (right). An example growth track is shown at the bottom-right to facilitate the readability of the axes. In all scenarios, protoplanets migrate several AU during gas accretion. This phenomenon is visually depicted as a significant horizontal displacement from the diagonal line $r_{\rm{core}}\,$$=\,$$r_{\rm{f}}$. The displacement is more pronounced in disks with high $\rm{St}$ (purple and blue groups) due to an earlier formation of the cores.}
        \label{fig:furthestCore_tsrf}%
        % Plot made by find_distantcore.py
    \end{figure*}

Here, we additionally varied the Stokes number (again 0.01, 0.03, and 0.06), the initial metallicity of $Z_{0}\,$$ =\,$$0.01$ and $0.02$, and the midplane turbulence of $\alpha_{\rm{t}}\,$$ =\,$$10^{-4}$ and $10^{-5}$ (see Table \ref{tab:simSet}). Placing protoplanets of $0.01\,M_{\oplus}$ at any location in the disk ($r_{0}\,$$\leq\,$$ R_{1}$), we find the location of the furthest cores for different formation times: $t_{0,\rm{p}}\,$$ =\,$$0.2, 0.3, 0.5$ and $0.7\,\mathrm{Myr}$. We performed the calculations for the disk sizes $R_{1}\,$$ =\,$$100$ and $300\,\mathrm{AU}$. We note that when $R_{1}\,$$ =\,$$300\,\mathrm{AU}$,  we obtain $\rm{St}_{\rm{dritft}}\,$$\approx \,$$0.02$ and $t_{0}\,$$\approx\,$$ 0.7\,\mathrm{Myr}$ for our fiducial values (see Fig.\,\ref{fig:St_drift}). For simplicity, we only varied a single parameter at a time and for the remaining disk parameters, we adhered to the values motivated in Sect.\,\ref{sec:pebbleAccretion}. However, we address this assumption in the discussion.

In Fig.\,\ref{fig:furthestCore}, we show the location of furthest cores for different scenarios (different bars) for different formation times of the protoplanet (different subfigures) and for different disk sizes (different color bars). We also include the location of the observed gaps in protoplanetary disks (horizontal green lines). First, we see that the location of the furthest core depends on the disk properties, including the disk size. Closer to the initial time, $t_{0}$, the pebble flux is higher and therefore the growth is faster when the protoplanet forms early on. However, the level of impact of varying the formation time will depend on the disk parameters. On the one hand, when $t_{0,\rm{p}}\,$$ =\,$$0.2\,\mathrm{Myr}$, the furthest cores form when $\rm{St}\,$$ =\,$$0.06$ (abbreviation "hst"). In these scenarios where there is a strong but short-lasting pebble flux, the furthest distance drops quickly when varying $t_{0,\rm{p}}$ (especially for the smaller disk); for example, for the case of $\rm{St}\,$$ =\,$$0.06$, $Z_{0}\,$$ =\,$$0.02$ ($\rm{hst.hz}$) and disk size $R_{1}\,$$ =\,$$100\,\mathrm{AU}$, initially the furthest core can form beyond $50\,\mathrm{AU}$, but if the protoplanets form $0.5\,\mathrm{Myr}$ later, the furthest core do not reach $10\,\mathrm{AU}$. On the contrary, in scenarios where $\rm{St}\,$$ =\,$$0.01$, the furthest cores are initially less distant but do not depend so strongly on the formation time due to the weak but longer-lasting pebble flux. Therefore, when varying $\rm{St}$, planet formation faces the dilemma that increasing the pebble flux will accelerate the decay and consequently requires protoplanets to start their growth earlier.

We also see that a higher $Z_{0}$ leads to the formation of more distant cores. Doubling $Z_{0}$ doubles the growth rate of protoplanets (see Eqs.\,\ref{eq:M2d}, \ref{eq:M3d}, and \ref{eq:Sigmap}) but does not affect the decay time of the flux, nor the migration rate. Therefore, increasing metallicity has a positive outcome on the formation of distant cores; for instance, when $t_{0,\rm{p}}\,$$ =\,$$0.2\,\mathrm{Myr}$, in the fiducial simulation ("fid" and "hz") the most distant core moves from $\sim$$ 30\,\mathrm{AU}$ to $\sim$$ 50\,\mathrm{AU}$ when $R_{1}\,$$ =\,$$100\,\mathrm{AU}$, and from $\sim$$ 70\,\mathrm{AU}$ to $\sim$$ 120\,\mathrm{AU}$ when $R_{1}\,$$ =\,$$300\,\mathrm{AU}$.

Decreasing the turbulence, $\alpha_{\rm{t}}$, increases the location of the most distant core. As the pebble scale height, $H_{\rm{p}}$, is smaller for lower turbulence, the 3D accretion rate is higher and the transition from 3D to 2D accretion occurs earlier (see Eq.\,\ref{eq:Racc_Hp}). Decreasing $\alpha_{\rm{t}}$ has a higher impact for the case where $\rm{St}\,$$ =\,$$0.01$; since the pebble scale-height is larger and the pebble flux is weaker, the 3D accretion stage is longer.

\section{Formation of wide-orbit gas giants}\label{sec:gasAccretion}
In this section, we include a simple model for gas accretion and study the planetary evolution after reaching $M_{\rm{iso}}$. In addition, we implement an alternative pathway for gas accretion that a planet can enter before reaching the pebble isolation mass but after the pebble accretion rate has decayed substantially.
\subsection{Simple prescription for gas accretion} \label{sec:gasAccretion_mini}

We assume that gas accretion starts with the contraction of the gaseous envelope at a rate suggested by \citet{Ikoma2000},
\begin{equation}\label{eq:Mdot_KH}
\dot{M}_{\mathrm{KH}}=10^{-5}\,M_{\oplus} \,\mathrm{yr}^{-1}\,\left(\frac{M}{10 \,M_{\oplus}}\right)^4\left(\frac{\kappa}{0.1 \mathrm{~m}^2 \mathrm{~kg}^{-1}}\right)^{-1}\,,
\end{equation}
where $\kappa$ is the opacity of the envelope. We take \mbox{$\kappa\,$$=\,$$0.005 \mathrm{~m}^2 \mathrm{~kg}^{-1}$} as in \citet{Johansen2019}. Since the contraction accelerates at higher mass, the planet might eventually become so massive that growth will be restricted by the supply of gas flowing into its Hill sphere. Once this occurs, the growth rate of the protoplanet will be equal to the rate of gas supply, as described by \citet{Tanigawa2016} and \citet{Ida2018}:
\begin{equation}\label{eq:Mdot_disk}
\dot{M}_{\mathrm{disk}}=0.29\left(\frac{H}{r}\right)^{-2}\left(\frac{M}{M_{\star}}\right)^{4 / 3} \Sigma_{\rm{g}}r^{2}\Omega \frac{\Sigma_{\mathrm{gap}}}{\Sigma_{\mathrm{g}}}\,,
\end{equation}
where $\Sigma_{\mathrm{gap}}/\Sigma_{\rm{g}}$ is the same as in Sect.\,\ref{sec:migration}.

The growth rate cannot be larger than the global accretion rate of the gas flux within the disk $\dot{\mathcal{M}}_{\rm{g}}$ from Eq.\,(\ref{eq:Mdot}).  Indeed, \citet{Lubow2006} estimated that the maximum accretion rate onto the protoplanet is approximately $80\%$ of the gas flux and therefore:
\begin{equation}\label{eq:Mdot_min}
\dot{M}=\min \left[\dot{M}_{\mathrm{KH}},\dot{M}_{\mathrm{disk}}, 0.8\lvert \dot{\mathcal{M}}_{\rm{g}}\lvert \right]\,.
\end{equation}

\subsection{Evolution of distant cores}

We first analyze the gas accretion applied to the furthest cores found in Fig.\,\ref{fig:furthestCore} of the set of simulations in Table $\ref{tab:simSet}$ when the protoplanet forms at $t_{0,\rm{p}}\,$$ =\,$$0.2\,\mathrm{Myr}$. In Fig.\,\ref{fig:furthestCore_tsrf}, we plot the core position vs the core formation time. We also display the final location at the end of the disk lifetime of $5\,\mathrm{Myr}$ of these cores after they have accreted gas. The figure shows clearly that when $\rm{St}\,$$ =\,$$0.03$ or $0.06$, the cores form at early epochs, causing the gas giants to migrate tens of AU before the disk lifetime. In the case of a large disk, the cores that form at around $\sim$$100\,\mathrm{AU}$ end up becoming gas giants close to $10\,\mathrm{AU}$. As these cores form when $\Sigma_{\rm{g}}$ is still high, which the migration rate scales linearly with (see Eq.\,\ref{eq:migI}), after they reach $M_{\rm{iso}}$ they still undergo very significant migration despite the gap-opening. When $\rm{St}\,$$ =\,$$0.01$, the core forms at later epochs, and consequently the protoplanets undergo less (but still significant) migration when accreting gas. Thus, since the most significant migration occurs within the first $\mathrm{Myr}$, these results remain consistent, even for a shorter disk lifetime of $3\,\mathrm{Myr}$.
  
The formation of wide-orbit gas giants further out than $10\,\mathrm{AU}$ is clearly very challenging: the high $M_{\rm{iso}}$ in the outer regions requires a high surface density of pebbles to grow and this can only occur in disks with strong but short-lasting pebble fluxes. Hence, the core forms at early epochs and undergoes a long migration path due to the high surface density of the gas.

According to direct imaging surveys, the occurrence of stars hosting at least one giant planet with a mass between $0.5$ and $14\,\mathrm{M_{\rm{Jup}}}$ orbiting at $20-300\,\mathrm{AU}$ is only around $1\%$ \citep{Vigan2017}. However, the current model we have developed does not adequately account for the formation of these distant gas giants. We go on to describe an alternative pathway for gas accretion below, which entails pebble flux decay and we investigate the feasibility of explaining wide-orbit gas giants via this pathway.

\subsection{The pebble decay pathway for gas accretion}

\begin{figure}
        %\centering
        \includegraphics[width=8.5cm]{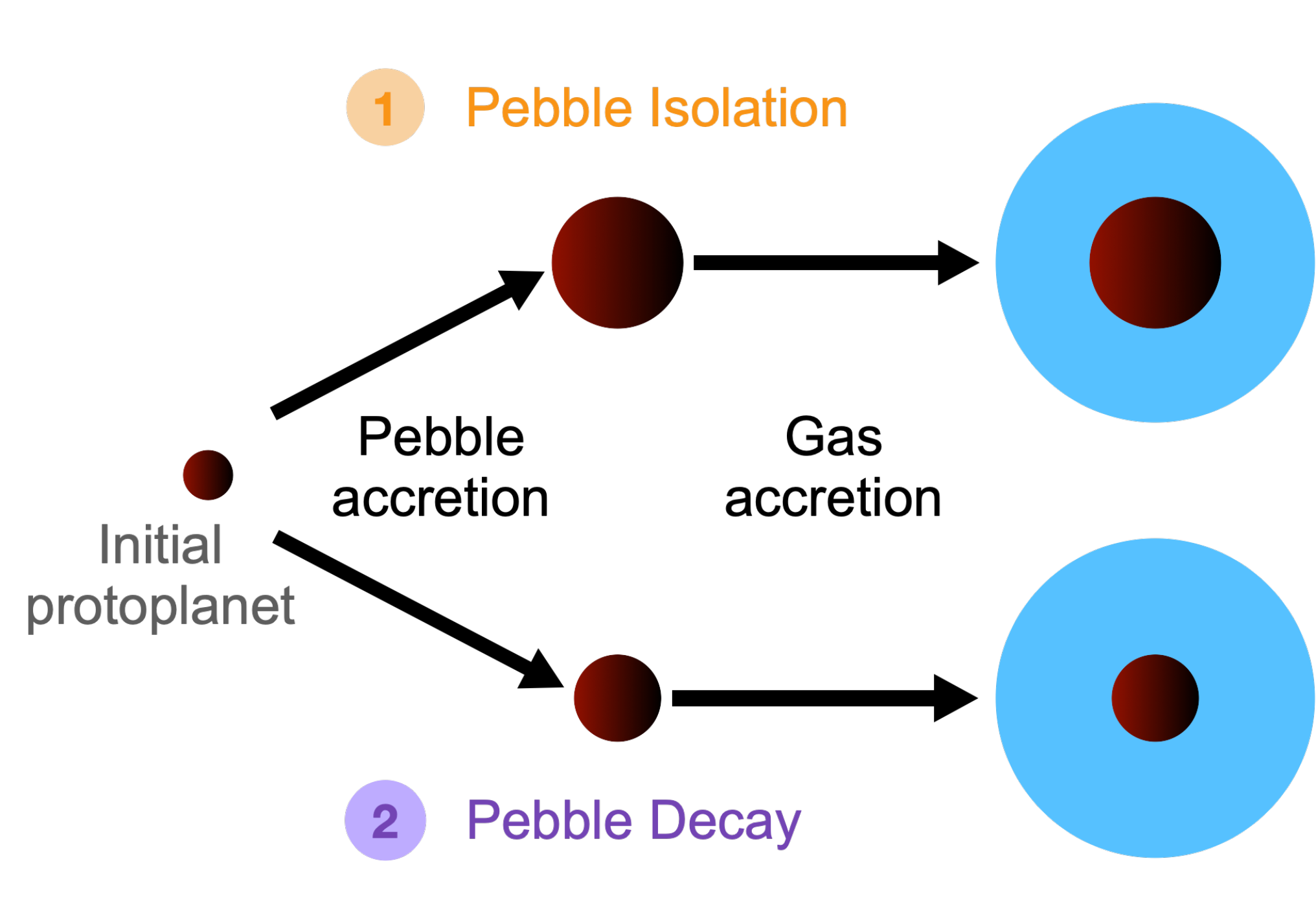}
        \caption{Two pathways for gas accretion based on the requirement that  pebble heating must cease for gas accretion to commence. 1) The pebble isolation pathway: the protoplanet reaches $M_{\rm{iso}}$ and the pebbles are trapped at the outer edge of the gap. 2) The pebble decay pathway: the pebble flux decays due to radial drift. If the protoplanet previously attained sufficient mass, its envelope contracts by radiative heat loss, resulting in gas accretion. The core mass distinguishes giant planets formed by the two pathways.}
         \label{fig:GasAccretionPaths}
   \end{figure}

The pebble isolation mass is very high in the outer regions; for example, to reach $M_{\rm{iso}}$ at $50\,\mathrm{AU}$ the protoplanet must form a core of $\sim\,$$ 50\,M_{\oplus}$. During pebble accretion, the impacting pebbles heat the atmosphere, preventing contraction and efficient gas accretion \citep{Lambrechts2014}. Hence, the conventional pathway for gas accretion is that it is necessary to reach $M_{\rm{iso}}$ to cool down the envelope and accrete gas. Nevertheless, when the pebble flux reduces with time due to the radial drift of pebbles, a protoplanet that did not reach $M_{\rm{iso}}$ can still accrete gas efficiently. We shall denote the two distinct formation channels as the "pebble isolation pathway" and the "pebble decay pathway," as illustrated in Fig.\,\ref{fig:GasAccretionPaths}.

   \begin{figure*}
    %\centering
    \includegraphics[width=18.3cm]{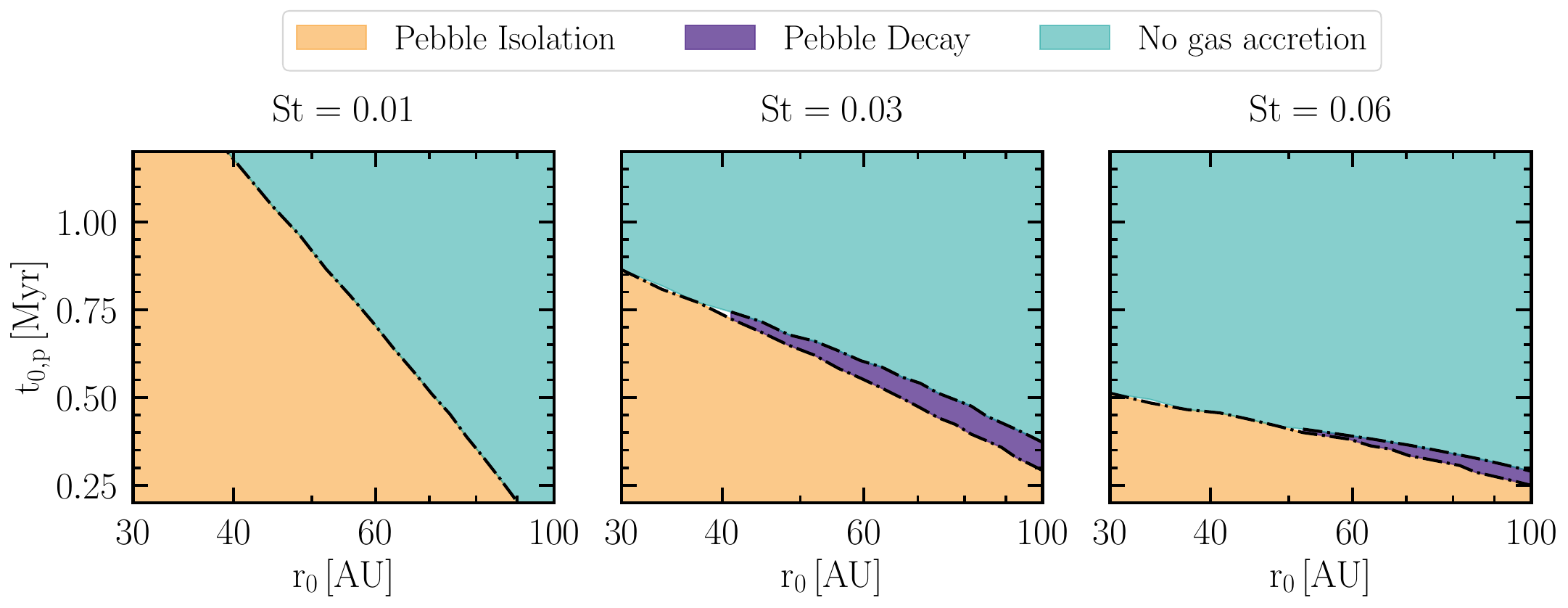}
    \caption{Gas accretion pathway depending on the initial position and formation time of a protoplanet of $0.01\,M_{\oplus}$ in disks with different pebble Stokes numbers. We employ a disk lifetime of $t_{\rm{f}}\,$$ =\,$$5\,\mathrm{Myr}$, an initial metallicity of $Z_{0}\,$$ =\,$$0.01$ and a midplane turbulence of $\alpha_{\rm{t}}\,$$ =\,$$10^{-4}$. The rest of the parameter values are listed in Table \ref{tab:fiducial}. \textit{Left}: Protoplanets cannot accrete gas via the pebble decay pathway when the pebbles are small and do not drift significantly. \textit{Center}: Under limited initial conditions, some gas giants form via the pebble decay pathway due to the short-lasting flux for a slightly higher value of $\rm{St}$. \textit{Right}: For even higher values of $\rm{St}$ the decay of the flux occurs earlier, and therefore, protoplanets must emerge at even earlier stages to reach $M_{\rm{iso}}$. The pebble decay pathway is less pronounced in this case.}
     \label{fig:initialConditions}
     % Plot done by find_Misoline
\end{figure*}

\begin{figure*}
    %\centering
    \includegraphics[width=18.3cm]{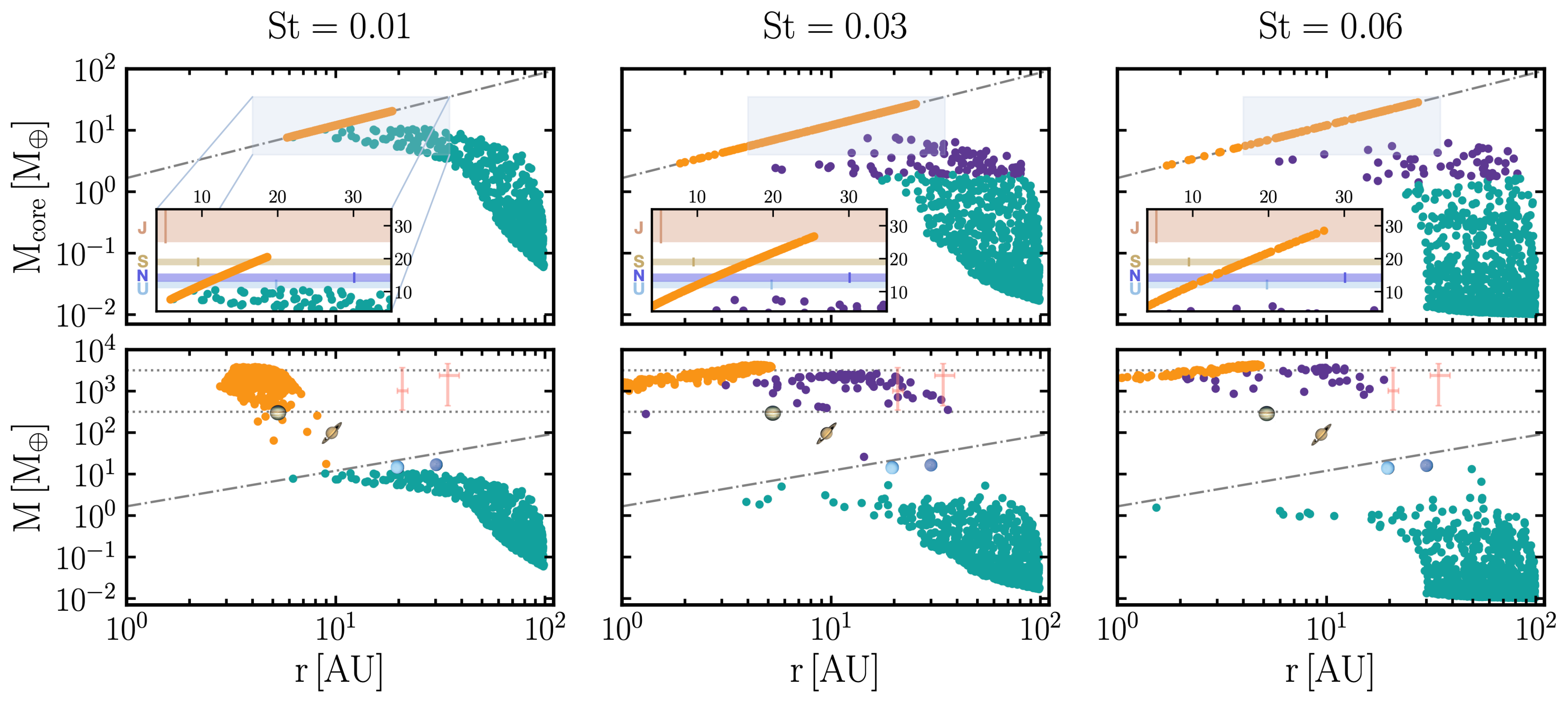}
    \caption{Population plot of core formation for $1000$ protoplanets randomly initialized from Fig.\,\ref{fig:initialConditions} (top). We indicate $M_{\rm{iso}}$ with gray dashed-dotted lines. We zoom in in the outer regions where the protoplanets reach $M_{\rm{iso}}$, and we indicate the metal content of the giant planets in the Solar System (see Sect.\,\ref{sec:SolarSystem} for discussion). A medium or high pebble Stokes number is required to form a planet with Jupiter's metallicity. Protoplanets that accrete gas via the pebble decay pathway (purple dots) acquire a significantly less massive core. Population plot including gas accretion for $t_{\rm{f}}\,$$ =\,$$5\,\mathrm{Myr}$ (bottom). The pebble isolation mass $M_{\rm{iso}}$, $1\,\mathrm{M_{Jup}}$ and $10\,\mathrm{M_{Jup}}$ are indicated as reference lines, as well as the giant planets in the Solar System and the protoplanets PDS 70 b and c are indicated with red error bars \citep[data from][]{Wang2021}. With a medium Stokes number and a short-lasting pebble flux, the formation of wide-orbit gas giants located up to $40\,\mathrm{AU}$ is possible due to the pebble decay pathway.}

     \label{fig:population plot}
     % Plot done by find_Misoline.py
\end{figure*}

We use  a simple description to implement the pebble decay pathway: if the time required for doubling the mass of a protoplanet via pebble accretion is longer than a certain threshold time, $\tau_{\rm{th}}$, pebble accretion stops and the protoplanet starts accreting gas. The mass-doubling timescale is
\begin{equation}\label{eq:newPathGas}
    \tau \approx \frac{M}{\dot{M}}\,,
\end{equation}
where $M$ is the mass of the protoplanet and $\dot{M}$ its growth rate via pebble accretion. When the pebble flux decays, $\dot{M}$ decreases and, consequently, $\tau$ will approach $\tau_{\rm{th}}$. Even small protoplanets that have experienced only limited growth can fulfill $\tau\,$$>\,$$\tau_{\rm{th}}$. However, when $M$ is very small, gas accretion is inefficient and the protoplanet will not grow within the lifetime of the protoplanetary disk.

The threshold time, $\tau_{\rm{th}}$, is a priori unknown and it will depend on the mass loading of heavy elements in the envelope, as it might take longer to cool down if the envelope is highly polluted \citep{Ikoma2000}. In \citet{Lambrechts2014}, these authors calculated the minimal accretion rates required to sustain a stable gas envelope and from their calculations, we infer that $\tau_{\rm{th}}$ is likely in the range between $10\,\mathrm{Myr}$ and $100\,\mathrm{Myr}$. We take $\tau_{\rm{th}}\,$$=\,$$10\,\mathrm{Myr}$ as the fiducial value.

We show in Fig.\,\ref{fig:initialConditions} the pathway that protoplanets would take depending on their initial position, $r_{0}$, and formation time, $t_{0,\rm{p}}$, for different Stokes numbers when the disk lifetime is extended up to $5\,\mathrm{Myr}$. For $\rm{St}\,$$ =\,$$0.03$ and $0.06$, a protoplanet initially placed at $30\,\mathrm{AU}$ needs to form earlier than $ 0.75\,\mathrm{Myr}$ and $0.5\,\mathrm{Myr,}$ respectively, to reach $M_{\rm{iso}}$. For $\rm{St}\,$$ =\,$$0.01$, the required formation time extends beyond $1.2\,\mathrm{Myr}$. However, a protoplanet placed at $100\,\mathrm{AU}$ will never reach $M_{\rm{iso}}$ when $\rm{St}\,$$=\,$$0.01$; furthermore, when $\rm{St}\,$$=\,$$0.03$ and $0.06$ the required formation times drop down to $0.3\,\mathrm{Myr}$ and $0.25\,\mathrm{Myr}$. Above the line that separates the protoplanets that reach $M_{\rm{iso}}$ and the ones that do not, in the $\rm{St}\,$$=\,$$0.03$ and $0.06$ scenarios there is a relatively narrow region where the body can efficiently accrete gas\footnote{We consider that a protoplanet undergoes "efficient gas accretion" when the protoplanet located at $r_{\rm{i}}$ has a higher mass than $M_{\rm{iso}}(r_{\rm{i}})$.} due to pebble flux decay.

Figure \ref{fig:population plot} displays the evolution of $1000$ protoplanets with initial conditions randomly chosen from the ranges shown in Fig.\,\ref{fig:initialConditions}. We compare the core masses with the total metal amount of the giant planets in the Solar System: Jupiter from $25$ to $45\,M_{\oplus}$ \citep{Wahl2017}, Saturn from $18$ to $20\,M_{\oplus}$ \citep{Mankovich2021}, and Uranus and Neptune from $11$ to $13\,M_{\oplus}$ and from $13$ to $15.5\,M_{\oplus}$ respectively \citep{Helled2011}. A discussion on Solar System formation is given in Sect.\,\ref{sec:SolarSystem}. The main takeaway from Fig.\,\ref{fig:population plot} is that considering two paths for gas accretion results in two types of gas giants. Gas giants formed via the pebble isolation pathway have a metal-rich core. As their core forms at early epochs, they undergo a strong migration and end up orbiting $\sim\,$$ 5\,\mathrm{AU}$ at the furthest. In contrast, the final position of gas giants formed via the pebble decay pathway can extend beyond $5\,\mathrm{AU}$ up to $40\,\mathrm{AU}$. They have a smaller core between $1.5$ and $8\,M_{\oplus}$. In addition, their formation is rather unusual, as the protoplanet must attain sufficient growth through pebble accretion to be able to accrete gas, while avoiding excessive migration and aligning with the decrease in pebble flux.
\section{Implications and limitations}\label{sec:implications}

In this section, we establish connections between our primary findings and the observations of protoplanetary disks and planetary systems. Additionally, we highlight the limitations that require further exploration in future studies. First, we delve into the evolutionary patterns concerning the overall solid mass of protoplanetary disks. Subsequently, we investigate the plausibility of planetary cores being responsible for the observed gaps within these disks. Moreover, we investigate the potential origin of the PDS 70 system, which encompasses two wide-orbit Jovian protoplanets, as well as the formation of wide-orbit gas giants observed in mature planetary systems. Lastly, we conclude by exploring the prospective scenarios of the formation of the Solar System.
\subsection{Solid mass evolution in protoplanetary disks}
Our analytical solution to the pebble flux problem allows us to also express  the temporal evolution of the total solid mass as:
\begin{equation}\label{eq:Ms}
        M_{\rm{s}}(t) =  \int_{0}^{\infty} 2\pi r\Sigma_{\rm{p}}(r, t)\,dr = M_{\rm{s,0}} T^{\,-\frac{1}{2(2-\gamma)}\left( 1 + \frac{2}{3}\frac{\chi\rm{St}}{\alpha}\right)}\,,
\end{equation}
where the initial solid-mass is $M_{\rm{s,0}}\,$$ =\,$$ M_{\rm{g,0}}\,$$\cdot\,$$ Z_{0}\,$$=\,$$\frac{2}{3}\frac{\dot{\mathcal{M}}_{g, 0}}{\nu_{1}}\frac{R^{2}_{1}}{(2-\gamma)}Z_{0}$, the pebble surface density, $\Sigma_{\rm{p}}$, is given by Eq.\,(\ref{eq:Sigmap}), and the rest of the parameters are specified in Sect.\,\ref{sec:pebbleAccretion}. This expression is valid for $t\,$$\ge\,$$t_{0}$, $t_{0}$ being the time at which the solids grow up to the fragmentation or drift limit at $R_{1}$. We ignore the dust mass loss in the earliest phases ($t\,$$<\,$$t_{0}$) and, therefore, a decrease in the initial metallicity $Z_{0}$ needs to be considered to apply properly to large disks of, for instance,\,$300\,\mathrm{AU}$.

\begin{figure}
    \centering
    \includegraphics[width=9cm]{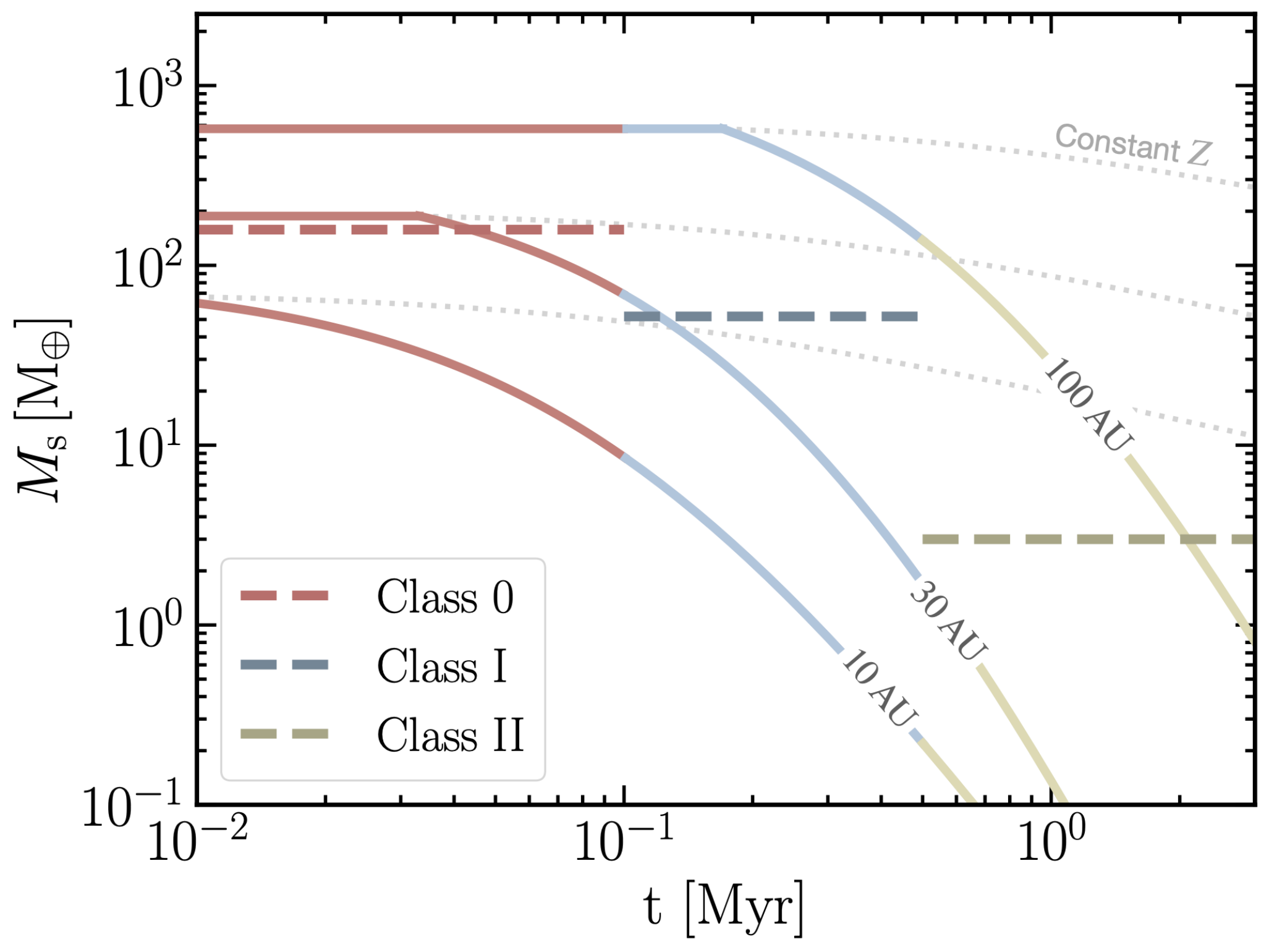}
    \caption{Comparison between the evolution of the solid mass reservoir according to Eq.\,(\ref{eq:Ms}) for different initial disk sizes $R_{1}$. We indicate the median dust masses derived by \citet{Tychoniec2020} for Class 0 and Class I sources in Perseus, and for Class II sources in Lupus by \citet{Ansdell2016}. The dashed lines indicate the median value for each evolutionary class. We assume that the evolutionary phases Class 0 and I last approximately $0.1\,\mathrm{Myr}$ and $0.5\,\mathrm{Myr,}$ respectively \citep{Dunham2014}. We assume that $Z_{0}\,$$=\,$$0.01$ remains constant until solids grow to the drift or fragmentation limit at $R_{1}$ as in Fig.\,\ref{fig:St_drift}. For $R_{1}\,$$=\,$$10, 30, 100\,\mathrm{AU}$, we estimate \mbox{$t_{0}\,$$=\,$$0.007, 0.03, 0.2 \,\mathrm{Myr}$} and $\rm{St}\,$$=\,$$0.02, 0.03, 0.03,$ respectively. Gray dotted lines indicate solid mass decrease solely due to gas accretion onto the star, i.e., \mbox{$M_{\rm{s}}(t)=Z_{0}M_{\rm{g}}(t)$}.}
     \label{fig:dustMass}
     % Plot done by DustMass.py
\end{figure}

Figure \ref{fig:dustMass} compares the solid mass evolution with the masses of Class 0 and I disks in the star-forming region Perseus derived by \citet{Tychoniec2020}, as well as for Class II sources in Lupus derived by \citet{Ansdell2016}. Their estimated median values for the dust mass in Class 0 and I disks are below the ones we calculated for a disk with $R_{1}\,$$=\,$$100\,\mathrm{AU}$. However, if we assume a more typical disk size of $R_{1}\,$$=\,$$30\,\mathrm{AU}$, the analytical expression lines up with the observed masses. This preliminary comparison is done only for Sun-like stars and the further exploration of parameter variations is left to future studies \citep[see also][]{Appelgren2023}.

\subsection{Considering whether distant core formation explain the observed gaps in disks}\label{sec:gaps}
The emergence of the observed substructures in the outer regions of protoplanetary disks has raised questions about their origin, since gas giants are rarely encountered in wide orbits in mature planetary systems \citep[see discussion in][]{Drazkowska2022}. Nevertheless, in all our scenarios, once the cores form in the outer regions via the pebble isolation pathway (see Fig.\,\ref{fig:GasAccretionPaths}), they migrate tens of AUs until the disk dissipates. That could explain why we observe multiple substructures beyond $10\,\mathrm{AU}$ that differ from an orbital location of $<\,$$10\,\mathrm{AU,}$ where most giant exoplanets are situated \citep[see Fig.\,1 in][]{Lodato2019}.

\citet{Bae2022} analyzed a sample of 62 protoplanetary disks observed at NIR and/or mm wavelengths containing rings and gaps, finding a maximum frequency of these substructures to occur at $20-50\,\mathrm{AU}$. However, some rings were also found further than $\sim$$100\,\mathrm{AU}$ from their central star (see their Fig.\,3(d)). When it comes to determining whether we can explain these substructures, we assume that protoplanets carve out gaps once they reach $M_{\rm{iso}}$ and neglect the plausible impact of fast migration regarding the gap-opening \citep{Kanagawa2020}. We hereby find that the furthest gap-forming cores form at distances from $20\,\mathrm{AU}$ to $80\,\mathrm{AU}$ in the most realistic scenarios (see Fig.\,\ref{fig:furthestCore}). We note that we only considered the $t_{0,\rm{p}}\,$$=\,$$0.7\,\mathrm{Myr}$ case for $R_{1}\,$$=\,$$300\,\mathrm{AU,}$ since it takes approximately $\sim\,$$ 0.7\,\mathrm{Myr}$ for the solids to grow to the drift-limit at $300\,\mathrm{AU}$. Overall, we find that the formation of distant cores is determined by the disk characteristics and formation of the initial protoplanet.

\textit{Disk parameters}: Our fiducial disk model of $Z_{0}\,$$=\,$$0.01$ and $R_{1}\,$$=\,$$100\,\mathrm{AU}$ contains an initial pebble mass reservoir of approximately $570\,M_{\oplus}$. In this case, we see gap formation as far out as $30\,\mathrm{AU}$. In order to form cores up to $50\,\mathrm{AU}$ or $80\,\mathrm{AU}$, we find necessary either higher metallicity of $Z_{0}\,$$=\,$$0.02$ or larger disk size of $R_{1}\,$$=\,$$300\,\mathrm{AU,}$ respectively, which both result in a higher solid mass reservoir \citep[see also previous study by][]{Ndugu2019}. In addition, we find that weak turbulence enhances distant core formation and that the outcome depends strongly on the pebble Stokes number. When there is a strong short-lasting flux, distant cores form only at the early epoch ($t\,$$<\,$$1\,\mathrm{Myr}$). On the contrary, less distant cores can form at a later epoch with a weak but long-lasting pebble flux instead (see Fig.\,\ref{fig:furthestCore_tsrf}). Protoplanets formed via the pebble decay pathway by a short-lasting pebble flux are also expected to carve out gaps at later stages after undergoing significant gas accretion (beyond $1\,\mathrm{Myr}$ or $2.5\,\mathrm{Myr}$ for $\rm{St}\,$$=\,$$0.06$ or $\rm{St}\,$$=\,$$0.03,$ respectively). Such gaps may exhibit less observational prominence, since they would form only after significant depletion of the solid mass.
    
\textit{Initial protoplanets}: We find that distant and rather early-formed initial protoplanets are required to form wide-orbit cores (see Fig.\,\ref{fig:initialConditions}). The formation time is further constrained if the initial protoplanet's mass is lower. These initial protoplanets could directly form by streaming instability if a local enrichment in metallicity of $Z\,$$\sim\,$$0.015$ and a high Stokes number up to at least $\rm{St}\,$$=\,$$0.01-0.1$ are satisfied \citep{Johansen2009, Bai2010}. A more recent study carried by \citet{Li&Youdin2021} found that the streaming instability can trigger planetesimal formation across a broader range of parameters, even extending to subsolar metallicities. Therefore, planetesimal formation might also take place in the outer regions. The water ice line, where the inward drifting pebbles sublimate, is a favorable location for fulfilling these criteria \citep[e.g.,][]{Ros&Johansen2013, Ida&Guillot2016, Drazkowska&Alibert2017, Schoonenberg2018}. If the ice lines of more volatile species (e.g., N$_{2}$, CO) are also favorable locations for forming planetesimals, this could explain the origin of distant protoplanets \citep{Qi2013}.

Finally, it has been suggested that envelope pollution due to the sublimation of accreted pebbles could decrease the critical metal mass for runaway gas accretion \citep[e.g.,][]{Lambrechts2014, Brouwers2020}. While the contraction of a polluted envelope could explain why ice giants like Uranus and Neptune were able to accrete a minor fraction of H$_{2}$/He below the pebble isolation mass, such a contraction among the polluted envelopes is not generally  relevant for understanding the formation of gas giants in our study.

\subsection{Explaining PDS 70 b and c and other wide-orbit gas giants }

According to our model, wide-orbit gas giants should be rare, since the protoplanet must attain sufficient growth through pebble accretion to be able to accrete gas while avoiding excessive migration and aligning with the decrease in pebble flux. With the inclusion of the pebble decay pathway for gas accretion, Fig.\,\ref{fig:population plot} demonstrates that it is possible to explain the existence of gas giants such as PDS 70 b, located at $20.8^{+1.3}_{-1.1}\,\mathrm{AU}$, and PDS 70 c, located at $34.3^{+4.6}_{-3.0}\,\mathrm{AU}$ \citep{Wang2021}. While the scenario of both protoplanets forming via the pebble decay pathway appears to be less common, it suggests the possibility that this pathway may have played a role \citep[see also][for an alternative formation channel in rings]{Jiang&Ormel}.

Our current study's main limitation when reproducing the PDS 70 system is that we consider individual protoplanets while, in reality, simultaneously growing bodies could interact with each other. Such interactions can lead to scattering processes that hinder their growth \citep{Levison2015, Bitsch2019}, or reduce material delivery to inner bodies due to the gap-opening of an outer protoplanet (\citealt{Weber2018}, but see also \citealt{Stammler2023}). In addition, having a pair of giant planets could slow down their joint migration \citep{Griveaud2023}. Further studies are required to understand how gravitational interaction between simultaneously evolving bodies impacts the viability of the pebble decay pathway.

Overall, we reproduced gas giants up to $40\,\mathrm{AU}$ in an initial disk size of $R_{1}\,$$=\,$$100\,\mathrm{AU}$. Subsequent studies may explore whether a larger disk and the inclusion of photoevaporation (which could slow down the late migration) can reproduce giant planets even further out or whether other mechanisms, such as formation in rings or gravitational instability, would have to be invoked to explain such systems as HR 8799 \citep{Marois2010}.

\subsection{Explaining the Solar System} \label{sec:SolarSystem}
In our simulations, most of the cores accrete solid material from the outer regions and end up as cold gas giants. This implies that the giant planets of the Solar System may have started their formation at distant locations, as suggested first by \citet{Bitsch2015b} and explored further by \citet{Pirani2019}. This distant formation is actually in line with the super-solar nitrogen abundance of Jupiter, since most of the nitrogen freezes outside of the N$_2$ snowline beyond $30\,\mathrm{AU}$ \citep{Bosman2019, Oberg2019}. In addition, the metal content of Jupiter is between $25\,M_{\oplus}$ and $45\,M_{\oplus}$ \citep{Wahl2017}. If most of the metal content is accreted during core formation, then according to Fig.\,\ref{fig:population plot}, the protoplanet must reach $M_{\rm{iso}}$ between $\sim$$ 24\,\mathrm{AU}$ and $46\,\mathrm{AU}$. Since modeling the formation of such massive cores growing solely via pebble accretion is challenging, late planetesimal accretion is usually deemed necessary to complete the total metal amount of Jupiter \citep[proposed by][]{Shiraishi2008}. However, \citet{Eriksson2022} demonstrated that during the gas accretion phase of the protoplanet, the accretion of planetesimals formed at planetary gap edges is a rather inefficient process. Therefore, an alternative mechanism for the late solid enrichment or the formation of a more massive core is necessary. In Fig.\,\ref{fig:furthestCore}, we could reproduce within specific scenarios cores as massive as Jupiter's via pebble accretion alone. According to Fig.\,\ref{fig:population plot}, however, most of these bodies would become more massive than Jupiter once the process of gas accretion is terminated. This might be due to our simplified gas accretion prescription. Effects that have not been considered in our model such as photoevaporation \citep[e.g.,][]{Rosotti2015} could also hamper the gas accretion process.

Regarding the formation of Saturn, the estimated metal content falls within the range of $18\,M_{\oplus}$ to $20\,M_{\oplus}$ \citep{Mankovich2021}, suggesting that it may have also reached $M_{\rm{iso}}$. It is plausible that a weak long-lasting pebble flux could account for the formation of gas giants even at distances of approximately $10\,\mathrm{AU}$ (see Fig.\,\ref{fig:population plot}). However, a stronger pebble flux is required to reach Jupiter's metal content. Consequently, a formation scenario of both Jupiter and Saturn can be described in a model with a long-lasting pebble flux with $\rm{St}\,$$=\,$$0.01$ and a higher solid mass reservoir with $Z_{0}\,$$=\,$$0.02$ (see Fig.\,\ref{fig:furthestCore}). Jupiter, being the first to form when the pebble flux was stronger, acquired a higher metal content and experienced more inward migration. On the other hand, Saturn could have gradually accreted from a less intense flux, ultimately settling further away. An alternative possibility is that Saturn formed through the pebble decay pathway, yet this seems less probable given its considerable metal content. Nevertheless, it is possible for protoplanets to accumulate additional metals through alternative mechanisms,  such as the accretion of gas enriched in volatiles deposited by drifting pebbles \citep{Schneider2021a, Schneider2021b}. Finally, for Uranus and Neptune, the most likely scenario is that they did not undergo efficient gas accretion, as they should have never reached $M_{\rm{iso}}$ \citep{Lambrechts2014}. This notion aligns well with the proposed formation of Jupiter and Saturn via a long-lasting pebble flux, as the lack of depletion would prevent Uranus and Neptune from accreting gas via the pebble decay pathway.

\section{Summary}\label{sec:sum}
In this paper, we report the discovery of a new analytical expression to describe the temporal decay of the pebble flux. We first derived two analytical forms of the evolution of the metallicity, provided in Eqs.\,(\ref{eq:Z_sol0}) or (\ref{eq:Z_sol1}), under the assumptions that, respectively, the pebble Stokes number $\rm{St}$ or $\rm{St}_{\chi}\,$$\equiv\,$$ \rm{St} \cdot \chi$ (where $\chi$ is the logarithmic gas pressure gradient) are constant both in time and space. We advocate to use constant $\rm{St}_{\chi}$ as this agrees best with numerical simulations that include both radial drift and fragmentation in limiting the dust size. We demonstrated that for particles with $\rm{St}\,$$\gtrsim\,$$ 0.01$, core growth rates are significantly overestimated when using a simplified model that assumes a constant pebble-to-gas ratio at all radii and times.

We then used our derived pebble flux model to study the formation of distant planetary cores that reach the pebble isolation mass and therefore are able to carve out gaps in the outer regions of protoplanetary disks. In a large disk of $100\,\mathrm{AU}$ in size, we found that a Moon-sized protoplanet, that emerges within the first $\sim$$0.5\,\mathrm{Myr}$ and is located beyond $50\,\mathrm{AU}$, can grow to become the core of a gas giant at $20-50\,\mathrm{AU}$ (see Fig.\,\ref{fig:furthestCore}). The most distant cores form when there is a strong, but short-lasting, pebble flux with a Stokes number of $\gtrsim\,$$ 0.03$ (see Sect.\,\ref{sec:distantcores}). An initial metallicity of $\gtrsim \,$$0.01$ and low turbulence of $\lesssim \,$$10^{-4}$ also promote the formation of distant cores. As these cores form early, they undergo a fast inward migration while they accrete gas, and by the end of the disk lifetime, they become giant planets orbiting at radii within $10\,\mathrm{AU}$. In larger disks (e.g., $R_{1}$$\,=$$\,300\,\mathrm{AU}$), cores more massive than $50\,M_{\oplus}$ could form beyond $50\,\mathrm{AU}$, but it is still a challenge to explain the formation of planetary cores beyond $80\,\mathrm{AU}$ (see Sect.\,\ref{sec:gaps}).

We also explored an alternative pathway for triggering gas accretion that we named the "pebble decay pathway" (see Fig.\,\ref{fig:GasAccretionPaths}). We have demonstrated that this pathway could explain the formation of wide-orbit gas giants such as PDS 70 b and c. This pathway is only possible when the pebble flux decays significantly before the disk lifetime, implying that pebbles must grow up to large sizes ($\rm{St}$$\,\gtrsim\,$$ 0.03$)  to undergo fast radial drift and deplete. Given the appropriate disk parameters and initial conditions of the protoplanets, the location of these gas giants ranges from $1\,\mathrm{AU}$ up to $40\,\mathrm{AU}$ (see Fig.\,\ref{fig:population plot}). According to our model, to form a wide-orbit gas giant the protoplanet must have attained sufficient mass by the time the pebble flux has decayed in order to accrete gas efficiently, but not be too massive to avoid excessive migration. Hence, these planets are rare, in agreement with the low occurrence of $\sim$$1\%$ of distant gas giants from direct imaging surveys. Since they never reach $M_{\rm{iso}}$, we predict a low bulk metallicity content for these wide-orbit gas giants.

\begin{acknowledgements}
    N.G. thanks Federico Finkel for his comments on the mathematical derivations. The authors also thank the anonymous referee for the comments that helped to improve the manuscript. A.J. is supported by the Swedish Research Council (Project grant 2018-04867), the Danish National Research Foundation (DNRF Chair grant DNRF159), and the Knut and Alice Wallenberg Foundation (Wallenberg Academy Fellow grant 2017.0287). A.J. further thanks the European Research Council (ERC Consolidator grant 724 687-PLANETESYS), the Göran Gustafsson Foundation for Research in Natural Sciences and Medicine, and the Wallenberg Foundation (Wallenberg Scholar KAW 2019.0442) for research support. M.L. acknowledges the ERC starting grant 101041466-EXODOSS.
\end{acknowledgements}

\bibliographystyle{aa_url}
\bibliography{biblio}

\begin{appendix} %First appendix

\section{Derivation of the analytical metallicity for a constant $\rm{St}$}\label{app:sol_0}

We assume that $\rm{St}$ is constant and that $Z(\tilde{r}, 1)$$\,=$$\,Z_{0}$. Substituting $\chi(\tilde{r}, T)$ from Eq.\,(\ref{eq:chi}), we get:
\begin{equation}
\begin{split}
    b(\tilde{r}, T) &= \frac{2}{3}\frac{\rm{St}}{\alpha}\left( \chi_{0} + (2-\gamma)\frac{\tilde{r}^{\,(2-\gamma)}}{T}\right)\\
    &= b_{0}\left( 1 + \frac{2-\gamma}{\chi_{0}}\frac{\tilde{r}^{\,(2-\gamma)}}{T}\right)\,,
\end{split}
\end{equation}
where $b_{0} $$\,=$$\, \frac{2}{3}\frac{\chi_{0}\rm{St}}{\alpha}$. The spatial derivative of $b(\tilde{r}, T)$ is:
\begin{equation}
    \frac{\partial b}{\partial \tilde{r}} = \frac{2}{3}\frac{\rm{St}}{\alpha}(2-\gamma)^{2}\frac{\tilde{r}^{\,(1-\gamma)}}{T} = b_{0} \frac{(2-\gamma)^{2}}{\chi_{0}}\frac{\tilde{r}^{\,(1-\gamma)}}{T}\,.
\end{equation}
Replacing $\frac{\partial b}{\partial \tilde{r}}$ in the general continuity Eq.\,(\ref{eq:general}) and multiplying the equation by the term $\left[ (2-\gamma)\,\tilde{r}^{\,(1-\gamma)}\right]^{-1}$, we get the governing equation:
\begin{equation}\label{eq:fulfil}
    \begin{split}
          & \frac{1}{2}\left[ \frac{1+b_{0}}{2-\gamma} \frac{1}{\tilde{r}^{(1-\gamma)}} + \left( \frac{b_{0}}{\chi_{0}} -2 \right)\frac{\tilde{r}}{T}\right] \frac{\partial Z}{\partial \tilde{r}} -(2-\gamma)\frac{\partial Z}{\partial T}\\
        =& \, \frac{Z}{T}\frac{b_{0}}{2}\left[ 1-\frac{2-\gamma}{\chi_{0}}\left(1 - \frac{\tilde{r}^{\,(2-\gamma)}}{T}\right)\right]
          \,.
    \end{split}
\end{equation}
This equation is a first-order linear PDE, which, in turn, can simply be expressed as:
\begin{equation}\label{eq:general_vanish}
   A(\tilde{r}, T) \frac{\partial Z}{\partial \tilde{r}} + B(\tilde{r}, T) \frac{\partial Z}{\partial T} = C(\tilde{r}, T, Z)\,.
\end{equation}
The corresponding Lagrange-Charpit system is
\begin{equation}\label{eq:Lagrange-Charpit}
    \frac{d\tilde{r}}{A(\tilde{r}, T)} = \frac{dT}{B(\tilde{r}, T)}\, = \frac{dZ}{C(\tilde{r}, T, Z)}.
\end{equation}
From the first equality, we get: 
\begin{equation}\label{eq:characteristic}
    \frac{d\tilde{r}}{\frac{1}{2}\left[ \frac{1+b_{0}}{2-\gamma} \frac{1}{\tilde{r}^{(1-\gamma)}} + \left( \frac{b_{0}}{\chi_{0}} -2 \right)\frac{\tilde{r}}{T}\right]} = -\frac{dT}{2-\gamma}\,.
\end{equation}
By making the change of variable $x$$\,=$$\,\tilde{r}^{(2-\gamma)}$,
\begin{equation}
    \frac{dx}{dT} = (2-\gamma)\tilde{r}^{1-\gamma}\frac{d\tilde{r}}{dT} = -\frac{1}{2}\left[ \frac{1 +  b_{0}}{2-\gamma} + \left( \frac{b_{0}}{\chi_{0}} -2 \right)\frac{x}{T}\right]\,.
\end{equation}
This equation is a homogeneous first-order ODE, meaning that it takes the form $ \frac{dx}{dT}$$\,=$$\,f\left(\frac{x}{T}\right)$ and therefore it can be solved by:
\begin{equation}\label{eq:homogeneous}
    \frac{dT}{T} = \frac{d u}{F(u) - u}\,,
\end{equation}
where $u$$\,=$$\,\frac{x}{T}$ and $F(u)$$\,=$$\,\frac{d x}{d T}$. Integrating the equation and substituting $x$$\,=$$\,\tilde{r}^{(2-\gamma)}$, we get the solution
\begin{equation}
    \ln{T} = -\frac{2\chi_{0}}{b_{0}}\ln{\left( \frac{1+b_{0}}{2(2-\gamma)} +\frac{b_{0}}{2\chi_{0}}\frac{\tilde{r}^{(2-\gamma)}}{T}\right)}+ p\,,
\end{equation}
where $p$ is an invariant. Rearranging the equation,
we have:\ \begin{equation}\label{eq:r}
    \tilde{r}(T) = \left[-\frac{\chi_{0}}{2-\gamma}\left(1 + \frac{1}{b_{0}} \right) T + p T^{-\frac{b_{0}}{2\chi_{0}} + 1}\right]^{\frac{1}{2-\gamma}}\,,
\end{equation}
or
\begin{equation}\label{eq:p}
    p = \left[\frac{\chi_{0}}{2-\gamma}\left(1 + \frac{1}{b_{0}} \right) + \frac{\tilde{r}^{(2-\gamma)}}{T}\right]T^{\frac{b_{0}}{2\chi_{0}}}\,.
\end{equation}
From the second equality in Eq.\,(\ref{eq:Lagrange-Charpit}), we get: 
\begin{equation}
    -\frac{dT}{2-\gamma}\, = \frac{dZ}{\frac{Z}{T}\frac{b_{0}}{2}\left[ 1-\frac{2-\gamma}{\chi_{0}}\left(1 - \frac{\tilde{r}^{\,(2-\gamma)}}{T}\right)\right]}\,.
\end{equation}
By replacing $\tilde{r}$ from Eq.\,(\ref{eq:r}) and rearranging the equation,
\begin{equation}
    \left(\frac{1}{2(2-\gamma)} + \frac{b_{0}}{2\chi_{0}}\right)\frac{dT}{T}-\frac{b_{0}}{2\chi_{0}}pT^{-\frac{b_{0}}{2\chi_{0}}-1}dT = \frac{dZ}{Z} \,.
\end{equation}
Integrating the ODE, we get:
\begin{equation}
    \ln{T^{\frac{1}{2(2-\gamma)}+ \frac{b_{0}}{2\chi_{0}}}} + p T^{-\frac{b_{0}}{2\chi_{0}}} + f(p)= \ln{Z}\,.
\end{equation}
By rearranging and replacing the invariant $p$ from Eq.\,(\ref{eq:p}), the general solution of the PDE is:
\begin{equation}
\begin{split}
    Z(\tilde{r}, T) =& T^{\frac{1}{2(2-\gamma)} + \frac{b_{0}}{2\chi_{0}}}\exp{\left[ \frac{\chi_{0}}{2-\gamma}\left( 1 + \frac{1}{b_{0}} \right)  + \frac{\tilde{r}^{(2-\gamma)}}{T}\right]}\\
    & \times f\left( \left[ \frac{\chi_{0}}{2-\gamma}\left( 1 + \frac{1}{b_{0}} \right)  + \frac{\tilde{r}^{(2-\gamma)}}{T}\right]T^{\frac{b_{0}}{2\chi_{0}}} \right)\,.
\end{split}
\end{equation}
To get the form of $Z(\tilde{r}, T)$ that fulfills the initial condition $Z(\tilde{r}, 1)$$\,=$$\,Z_{0}$, since we have:
\begin{equation}
\begin{split}
    Z(\tilde{r}, 1) =& \exp{\left[ \frac{\chi_{0}}{2-\gamma}\left( 1 + \frac{1}{b_{0}} \right)  + \tilde{r}^{(2-\gamma)}\right]}\\
    & \times f\left( \left[ \frac{\chi_{0}}{2-\gamma}\left( 1 + \frac{1}{b_{0}} \right)  + \tilde{r}^{(2-\gamma)}\right] \right) = Z_{0}\,,
\end{split}
\end{equation}
$f(p)$ must take the form $f(p)$$\,=$$\,Z_{0}e^{-p}$. Setting everything together, we obtain the solution:
\begin{spacing}{0.3}
\begin{equation}\label{eq:Z_sol0_app}
\begin{split}
    Z(\tilde{r}, T) =&\;Z_{0}T^{\frac{1}{2(2-\gamma)} + \frac{ b_{0}}{2\chi_{0}}} \\
    &\times \exp\biggl\{ {-\left[\frac{\chi_{0}}{2-\gamma}\left(1 + \frac{1}{b_{0}}\right) + \frac{\tilde{r}^{(2-\gamma)}}{T}\right]\cdot\left[T^{\frac{b_{0}}{2\chi_{0}}}-1\right]}\biggl\}\,.
\end{split}
\end{equation}
\end{spacing}
\noindent By replacing $b_{0}$, we have:\ 
\begin{equation}
    \begin{split}
    Z(\tilde{r}, T) = & Z_{0}T^{\frac{1}{2(2-\gamma)} + \frac{ \rm{St}}{3\alpha}}  \\
    & \times \exp\biggl\{ {-\left[\frac{1}{2(2-\gamma)}\left(2\chi_{0} + \frac{3\alpha}{\rm{St}}\right) + \frac{\tilde{r}^{(2-\gamma)}}{T}\right]\cdot \left[T^{\frac{\rm{St}}{3\alpha}}-1\right]}\biggl\}\,.
    \end{split}
\end{equation}
\section{Derivation of the analytical metallicity for a constant ${\rm St}_\chi$}\label{app:sol_1}
Here, we follow the same procedure as in previous appendix. Multiplying  the general equation by the term $\left[ (2-\gamma)\,\tilde{r}^{\,(1-\gamma)}\right]^{-1}$ we get:\ 
\begin{equation}
        \frac{1}{2}\left[  \frac{1+b_{0}}{2-\gamma} \frac{1}{\tilde{r}^{(1-\gamma)}} -2\frac{\tilde{r}}{T}\right] \frac{\partial Z}{\partial \tilde{r}} -(2-\gamma) \frac{\partial Z}{\partial T}
= \, \frac{Z}{T} \frac{b_{0}}{2} \,.
\end{equation}
The corresponding Lagrange-Charpit system is:
\begin{equation}\label{eq:Lagrange-Charpit2}
    \frac{d\tilde{r}}{\frac{1}{2}\left[  \frac{1+b_{0}}{2-\gamma} \frac{1}{\tilde{r}^{(1-\gamma)}} -2\frac{\tilde{r}}{T}\right] } = - \frac{dT}{2-\gamma}\, = \frac{dZ}{\frac{Z}{T} \frac{b_{0}}{2} }.
\end{equation}
The second equality can be directly integrated to get:
\begin{equation}
    Z(\tilde{r}, T) = T^{-\frac{b_{0}}{2(2-\gamma)}}f(p)\,.
\end{equation}
We can rewrite the first equality from Eq.\,(\ref{eq:Lagrange-Charpit2}) as:
\begin{equation}\label{eq:linear2}
    \frac{d\tilde{r}}{dT} = -\frac{1}{2(2-\gamma)} \left[ \frac{1+b_{0}}{2-\gamma} \frac{1}{\tilde{r}^{(1-\gamma)}} - 2\frac{\tilde{r}}{T}\right]\,.
\end{equation}
Making again the change in the variable $x $$\,=$$\, \tilde{r}^{(2-\gamma)}$,
\begin{equation}
 \frac{dx}{dT} = (2-\gamma)\tilde{r}^{(1-\gamma)} \frac{d\tilde{r}}{dT} = -\frac{1}{2} \left[ \frac{1+b_{0}}{2-\gamma} - 2\frac{x}{T}\right]\,,
\end{equation}
which is also an homogeneous first-order ODE that can be solved as described in Eq.\,(\ref{eq:homogeneous}). The solution is then:
\begin{equation}
    p = \frac{1}{2-\gamma}\ln{T} + \frac{2}{1+b_{0}}\frac{\tilde{r}^{(2-\gamma)}}{T}\,.
\end{equation}
The general solution for the PDE is therefore:\ \begin{equation}
    Z(\tilde{r}, T) = T^{-\frac{b_{0}}{2(2-\gamma)}}f\left(\frac{1}{2-\gamma}\ln{T} + \frac{2}{1+b_{0}}\frac{\tilde{r}^{(2-\gamma)}}{T} \right)\,.
\end{equation}
To fulfill the initial condition $Z(\tilde{r}, T)$$\,=$$\,Z_{0}$, since
\begin{equation}
    Z(\tilde{r}, 1) = f\left(\frac{2}{1+b_{0}}\tilde{r}^{(2-\gamma)}\right)= Z_{0}\,,
\end{equation}
 $f(p)$ must take the form of $f(p)$$\,=$$\,Z_{0}$. Hence, the solution is
\begin{equation}
    Z(\tilde{r}, T) = Z_{0} T^{-\frac{b_{0}}{2(2-\gamma)}}\,,
\end{equation}
and by replacing $b_{0} $$\,=$$\, \frac{2}{3}\frac{\chi\cdot\rm{St}}{\alpha}$, we get
\begin{equation}
    Z(\tilde{r}, T) = Z_{0} T^{-\frac{1}{2-\gamma}\frac{\chi\cdot\rm{St}}{3\alpha}}\,.
\end{equation}

\section{An alternative method for determining the evolving metallicity}\label{app:modelZ_diff}

We describe an alternative way of deriving the evolution of the metallicity. First, we use simplified expressions for the gas surface density and gas flux, commonly used to describe the inner regions,
\begin{spacing}{0.3}
\begin{equation}\label{eq:flux}
    \Sigma_{\rm{g}}(r, t)= \frac{\dot{\mathcal{M}}_{g, 0}}{3\pi \nu_{1}\tilde{r}^{\gamma}}T^{-\frac{5/2-\gamma}{2-\gamma}}\,,
\end{equation}
\end{spacing}
\begin{equation}
    \dot{\mathcal{M}}_{g}(t)= \dot{\mathcal{M}}_{g, 0}T^{-\frac{5/2-\gamma}{2-\gamma}}\,.
\end{equation}
Recalculating $\frac{\partial \ln P}{\partial \ln r}$ with the simplified $\Sigma_{\rm{g}}(r, t)$ (see Eq.\,\ref{eq:chi}), we get that $\chi $$\,=$$\,\chi_{0}$. The ratio between the radial velocities from Eqs.\,(\ref{eq:vgas}) and (\ref{eq:vp}) is:
\begin{equation}\label{eq:v_ratio}
    \frac{v_{\rm{r, p}}}{v_{\rm{r, g}}}  = \frac{1 + \frac{2}{3}\frac{\chi\cdot \rm{St}}{\alpha}}{1+\mathrm{St}^2}\,,
\end{equation}
From Eq.\,(\ref{eq:Mdisk}), we know that the total mass of the gas disk is:
\begin{equation}\label{eq:Mass}
    M_{g}(t) = \frac{2}{3}\frac{\dot{\mathcal{M}}_{g, 0}}{\nu_{1}}\frac{R^{2}_{1}}{(2-\gamma)}T^{-\frac{5/2-\gamma}{2-\gamma} + 1}=2(2-\gamma) t_{\rm{s}}\dot{\mathcal{M}}_{g}(t)T\,.
\end{equation}
where $t_{\rm{s}}$ is the viscous timescale from Eq.\,(\ref{eq:ts}). We derive the mass over time as:
\begin{equation}\label{eq:nose}
\begin{split}
    \frac{dM_{\rm{g}}}{d t}&=2(2-\gamma)t_{\rm{s}}\dot{\mathcal{M}}_{g, 0}\frac{(-1)}{2(2-\gamma)t_{\rm{s}}}T^{-\frac{5/2-\gamma}{2-\gamma}} \\
    &= -\dot{\mathcal{M}}_{g, 0}T^{-\frac{5/2-\gamma}{2-\gamma}} = - \dot{\mathcal{M}}_{g}(t)\,.
\end{split}
\end{equation}
Assuming that $Z(r, t)$$\,\approx$$\, Z(t)$, we can relate the mass of the gas and the mass of the pebbles:
\begin{equation}
\begin{split}
M_{\rm{p}}(r, t) &= \int 2\pi r\Sigma_{\rm{p}}(r, t) d r \\
&\approx Z(t) \int 2\pi r \Sigma_{\rm{g}}(r, t) d r = Z(t)M_{\rm{g}} (t)\,.
\end{split}
\end{equation}
Therefore, $M_{\rm{p}}(t)$$\,\approx$$\, Z(t)M_{\rm{g}}(t)$. Taking the derivative of this equation over time, we have:\ 
\begin{equation}
    \frac{d M_{\rm{p}}(t)}{d t} = \frac{d Z}{d t} M_{\rm{g}}(t) + \frac{d M_{\rm{g}}}{dt}Z(t).
\end{equation}
As $\frac{dM_{\rm{g}}}{d t} $$\,=$$\, -\dot{\mathcal{M}}_{g}(t)$ (see Eq.\,\ref{eq:nose}), we assume that $\frac{dM_{\rm{p}}}{d t} $$\,=$$\, -\dot{\mathcal{M}}_{p}(t)$ as well. Hence,
\begin{spacing}{0.3}
\begin{equation}
    -\dot{\mathcal{M}}_{p}(t) = \frac{d Z}{d t} M_{\rm{g}}(t) -\dot{\mathcal{M}}_{g}(t)Z(t)\,,
\end{equation}
\end{spacing}
\begin{equation} \label{eq:tosolve}
    Z(t)\dot{\mathcal{M}}_{g}(t)\,\left(1- \frac{v_{\rm{r, p}}}{v_{\rm{r, g}}}\right) = \frac{d Z}{d t} M_{\rm{g}}(t)\,.
\end{equation}
Replacing Eq.\,(\ref{eq:Mass}),
\begin{spacing}{0.4}
\begin{equation}
    \left(1- \frac{v_{\rm{r, p}}}{v_{\rm{r, g}}}\right) \frac{1}{2(2-\gamma)t_{\rm{s}}}\frac{1}{T}d t= \frac{d Z}{Z}\,,
\end{equation}
\end{spacing}
\begin{equation}\label{eq:third}
    Z(t)= Z_{0}T^{\frac{1}{2(2-\gamma)}\left(1- \frac{v_{\rm{r, p}}}{v_{\rm{r, g}}}\right)}\,.
\end{equation}
Assuming that $\frac{1}{\rm{St}^{2} + 1}\,$$\approx\,$$1$, from Eq.\,(\ref{eq:v_ratio}), we have:\ 
\begin{equation}
    Z(t)= Z_{0}T^{-\frac{1}{2-\gamma}\frac{\chi\cdot\rm{St}}{3\alpha}}\,,
\end{equation}
and we get the same result as that of Eq.\,(\ref{eq:Z_sol1}).

\section{Attempt to find the metallicity for a nonconstant $\rm{St}$}\label{app:St_nonconstant}

We show why it does not appear possible to solve the PDE for nonconstant $\rm{St}$. First, we note that to date, there has been no analytical expression of $\mathrm{St}(r, t)$ available for describing the growth and transport simultaneously \citep[see][for analytical expression for growth]{Drazkowska2021}. We could, however, use the expression $\mathrm{St}(r)$$\,\approx$$\, \rm{St}_{\rm{drift}}$ or $\rm{St}_{\rm{frag}}$ for outer regions and assume that particles reach the growth limit at approximately the same time at all locations. First, we replace $\mathrm{St}(r)$$\,=$$\,\rm{St}_{\rm{drift}}$ in Eq.\,(\ref{eq:b}) and substitute $\gamma$$\,=$$\,1$ for simplicity. We rewrite $b$ and its dimensionless spatial derivative as:
\begin{equation}
    b = b_{\rm{s}} \tilde{r}^{-\frac{1}{2}}\,, \quad \quad \frac{\partial b}{\partial \tilde{r}} = -\frac{b_{\rm{s}}}{2}\tilde{r}^{-\frac{3}{2}}\,.
\end{equation}
Replacing them in the PDE from Eq.\,(\ref{eq:general}) with $\gamma$$\,=$$\,1$,
\begin{equation}
         \frac{1}{2}\left[ 1 + b_{\rm{s}}\tilde{r}^{-\frac{1}{2}} -2\frac{\tilde{r}}{T}\right] \frac{\partial Z}{\partial \tilde{r}} -\frac{\partial Z}{\partial T}  = \frac{Z}{T}  \frac{b_{\rm{s}}}{2}\tilde{r}^{-\frac{1}{2}}\left[  1 + \frac{T}{2\tilde{r}}\right].
\end{equation}
This equation is a first-order linear PDE, and the characteristic equation is:
\begin{equation}\label{eq:characteristic3}
    \frac{d\tilde{r}}{\frac{1}{2}\left[ 1 +  b_{\rm{s}}\tilde{r}^{-\frac{1}{2}} -2 \frac{\tilde{r}}{T}\right]} = -dT\,.
\end{equation}
Contrary to the linear Eqs.\,(\ref{eq:characteristic}) and (\ref{eq:linear2}) for constant $\rm{St}$ and $\rm{St}_{\chi}$ respectively, Eq.\,(\ref{eq:characteristic3}) is nonlinear. When substituting \mbox{$\rm{St}$$\,=$$\,\rm{St}_{\rm{frag}}$} from Eq.\,(\ref{eq:St_fr}), the characteristic equation is also nonlinear. Due to the nonlinearity, there is no straightforward way to an analytical solution that we are aware of.

\section{Comparison of analytical pebble flux models}\label{app:comparison_dotMp}
\begin{figure}
    \centering
    \includegraphics[width=9cm]{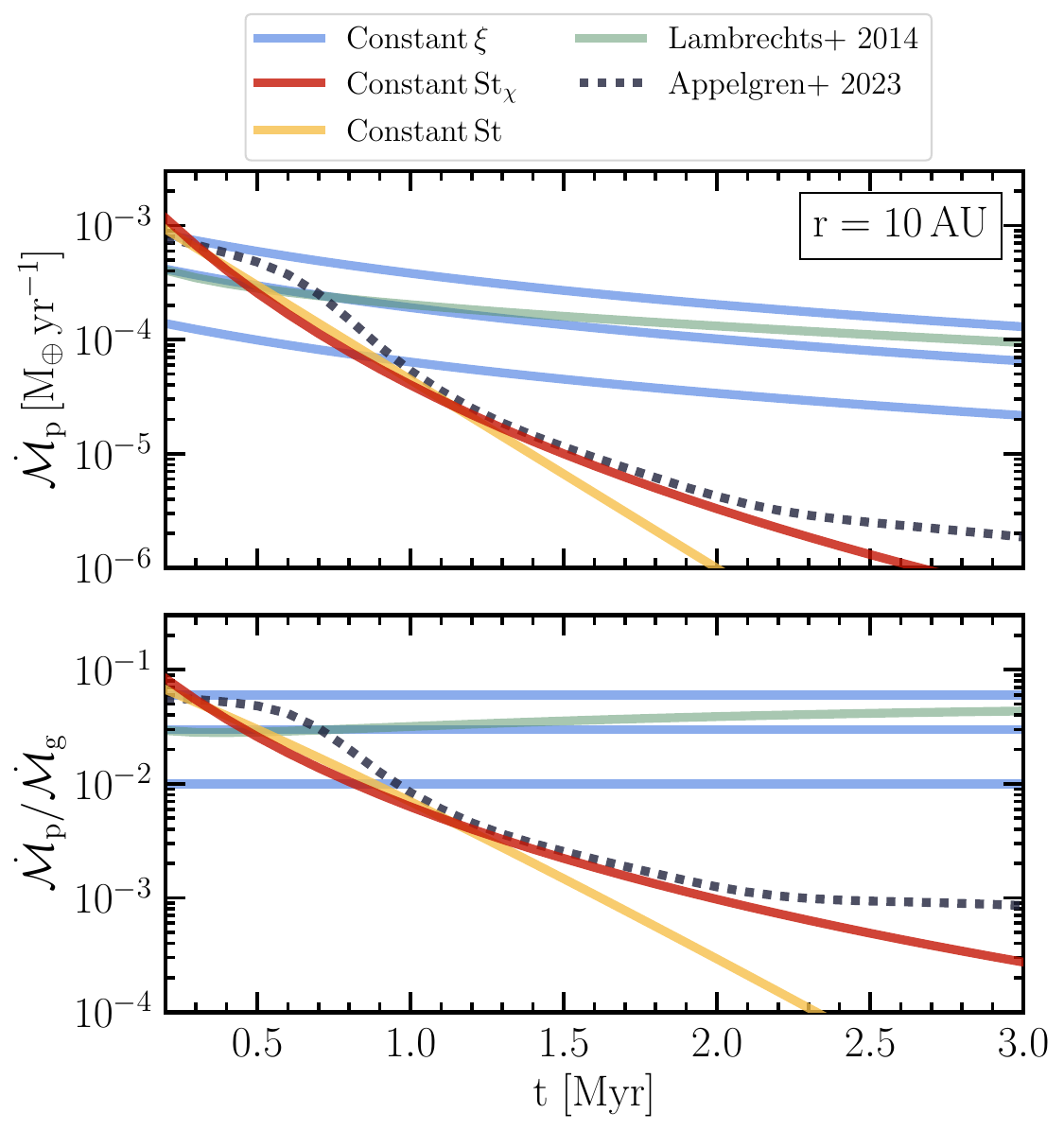}
    \caption{Evolution of the pebble flux (top) and the pebble-to-gas flux ratio (bottom) at $10\,\mathrm{AU}$ according to different analytical models and according to the simulation from \citet{Appelgren2023}. Analytical models are described in Eq.\,(\ref{eq:xi}) for a constant $\xi$$\,=$$\,0.01, 0.03$ or $0.06$, Eqs.\,(\ref{eq:Z_sol1}) and (\ref{eq:Z_sol0}) for a constant $\rm{St}_{\chi}$ and $\rm{St,}$ respectively, and Eq.\,(\ref{eq:michiel}) for the pebble flux derived by \citet{Lambrechts&Johansen2014}. We assumed $Z_{0}$$\,=$$\,0.008$ and $\dot{\mathcal{M}}_{\rm{g,0}}$$\,=$$\,6\times 10^{-8}\,M_{\odot}\,\mathrm{yr^{-1}}$ to match the simulation, and the rest of the parameter values are listed in Table \ref{tab:fiducial}. The simulation shows that, for the given disk parameters, the pebbles deplete on a shorter timescale than the gas due to the radial drift. This behavior is exhibited in the analytical models with a constant $\rm{St}_{\chi}$ and constant $\rm{St}$. }
     \label{fig:pebbleFluxes}
     % Plot done by model_testing.py
\end{figure}

We compare our new analytical models for the pebble flux with existing analytical models from the literature. Similarly to previous studies \citep[e.g.,][]{Johansen2019, Liu2019}, we consider a tightly coupled evolution of solids and gas, leading to the calculation of the pebble flux using the equation:
\begin{equation}\label{eq:xi}
    \dot{\mathcal{M}}_{\rm{p}} = \xi \dot{\mathcal{M}}_{\rm{g}}\,,
\end{equation}
 where $\xi$ represents a constant parameter. Additionally, we include a comparison with the model proposed by \citet{Lambrechts&Johansen2014}, which assumes a pebble formation front. The expression is given by:
\begin{equation}\label{eq:michiel}
\begin{split}
    \dot{\mathcal{M}}_{\rm{p}} \approx&\, 9.5\times 10^{-5}\left( \frac{\beta}{500\,\mathrm{g\,cm^{-2}}}\right)\left( \frac{M_{\star}}{M_{\odot}}\right)^{1/3}\\
    & \times \left( \frac{Z_{0}}{0.01}\right)^{5/3} \left( \frac{t}{\mathrm{Myr}}\right)^{-1/3} \,M_{\oplus}\,\mathrm{yr^{-1}}\,,
\end{split}
\end{equation}
where $\beta$ is
\begin{equation}
    \beta=\frac{\dot{\mathcal{M}}_{\rm{g,0}}}{3\pi \nu_{1}}\left(\frac{R_{1}}{\mathrm{AU}}\right)^{\gamma}\exp{\left(-\frac{t}{t_{\rm{f}}}\right)}\,.
\end{equation}
All the parameter values are specified in Table \ref{tab:fiducial}. Through a comparative analysis with the numerically calculated flux by \citet{Appelgren2023}, we demonstrate in Fig.\,\ref{fig:pebbleFluxes} that our new analytical pebble flux models exhibit a greater similarity to the numerical results than previous analytical approaches.

\end{appendix}

\end{document}